\documentstyle[12pt,epsfig,color,rawfonts,pictex]{article}
\newcommand{\Z}{{\sf Z \!\!\! Z}}

\newcommand{\1}{{\sf 1 \!\! 1}}
\newcommand{\0}{{\sf 0 \!\! 0}}
\newcommand{\p}{\partial}

\newcommand{\kbra}[2]{|#1\rangle \langle #2|}
\makeatletter
\@addtoreset{equation}{section}
\makeatother

\title{Exceptional Confinement in $G(2)$ Gauge Theory}

\author{K. Holland$^{{\rm a}}$, P. Minkowski$^{{\rm b}}$, M. Pepe$^{{\rm b}}$ 
and U.-J. Wiese$^{{\rm b}}$\footnote{on leave from MIT}
\\ \\
$^{{\rm a}}$ Department of Physics, University of California at San Diego \\
La Jolla, CA 92093, U.S.A. \\ \\
$^{{\rm b}}$ Institute for Theoretical Physics \\
Bern University, Sidlerstrasse 5, CH-3012 Bern, Switzerland \\}

\begin{document}

\maketitle

\vspace{-1cm}

\begin{abstract} \normalsize

We study theories with the exceptional gauge group $G(2)$. The 14 adjoint 
``gluons'' of a $G(2)$ gauge theory 
transform as $\{3\}$, $\{\overline 3\}$ and $\{8\}$ under the subgroup $SU(3)$,
and hence have the color quantum numbers of ordinary quarks, anti-quarks and 
gluons in QCD. Since $G(2)$ has a trivial center, a ``quark'' in the $\{7\}$ 
representation of $G(2)$ can be screened by ``gluons''. As a result, in $G(2)$ 
Yang-Mills theory the string between a pair of static ``quarks'' can break. In 
$G(2)$ QCD there is a hybrid consisting of one ``quark'' and three ``gluons''.
In supersymmetric $G(2)$ Yang-Mills theory with a $\{14\}$ Majorana ``gluino''
the chiral symmetry is $\Z(4)_\chi$. Chiral symmetry breaking gives rise to 
distinct confined phases separated by confined-confined domain walls. A scalar 
Higgs field in the $\{7\}$ representation breaks $G(2)$ to $SU(3)$ and allows 
us to interpolate between theories with exceptional and ordinary confinement. 
We also present strong coupling lattice calculations that reveal basic features
of $G(2)$ confinement. Just as in QCD, where dynamical quarks break the $\Z(3)$
symmetry explicitly, $G(2)$ gauge theories confine even without a center. 
However, there is not necessarily a deconfinement phase transition at finite 
temperature.

\end{abstract}
\newpage 
\section{Introduction}

Understanding confinement and the dynamical mechanism behind it is a big 
challenge in strong interaction physics. In $SU(3)$ Yang-Mills theory
confinement is associated with the $\Z(3)$ center of the gauge group. Since the
center symmetry is unbroken at low temperatures, an unbreakable string confines
static quarks in the fundamental $\{3\}$ representation to static anti-quarks 
in the $\{\overline 3\}$ representation. In the high-temperature deconfined 
phase the Polyakov loop \cite{Pol78,Sus79} gets a non-zero expectation value 
and the $\Z(3)$ symmetry breaks spontaneously. As a result, there are three 
distinct deconfined phases. Potential universal behavior at the deconfinement
phase transition is described by an effective 3-d scalar field theory for the
Polyakov loop \cite{Sve82}. The center symmetry and its spontaneous
breakdown were recently reviewed in \cite{Hol01,Gre03}. In full QCD the $\Z(3)$
symmetry is explicitly broken because quarks transform non-trivially under the 
center. As a result, the string connecting a quark and an anti-quark can break 
via the creation of dynamical quark-anti-quark pairs. Still, color remains 
confined and non-Abelian charged states --- like single quarks or gluons --- 
cannot exist. 

In this article we ask how confinement arises in a theory whose gauge group has
a trivial center. The simplest group with this property is $SO(3) = 
SU(2)/\Z(2)$. While $SO(3)$ has a trivial center, its universal covering group 
$SU(2)$ has the non-trivial center $\Z(2)$. Similarly, $SU(N_c)/\Z(N_c)$ has a 
trivial center and the corresponding universal covering group $SU(N_c)$ has the
non-trivial center $\Z(N_c)$. When one formulates Yang-Mills theories on the 
lattice, one usually works with Wilson parallel transporters in the universal 
covering group $SU(N_c)$. However, one can also work with parallel transporters
taking values in the group $SU(N_c)/\Z(N_c)$. In that case, it is impossible to
probe the gluon theory with static test quarks represented by Polyakov loops in
the fundamental representation of the gauge group. Instead one is limited to 
purely gluonic observables. In fact, $SO(3) = SU(2)/\Z(2)$ gauge theories have 
been studied in detail on the lattice 
\cite{Gre81,Bha81,Hal81,Can82,Che96,Dat98,deF03}. One finds that lattice 
artifacts --- namely center monopoles --- make it difficult to approach the 
continuum limit in this formulation. There is a phase transition in which the 
lattice theory sheds off these artifacts, and one then expects it to be 
equivalent to the standard $SU(2)$ Yang-Mills theory in the continuum limit. 
This suggests that it is best to formulate lattice gauge theories using the 
universal covering group, e.g. $SU(N_c)$ rather than $SU(N_c)/\Z(N_c)$, in 
order to avoid these lattice artifacts. The universal covering group of $SO(N)$
is $Spin(N)$ which
also has a non-trivial center: $\Z(2)$ for odd $N$, $\Z(2) \otimes \Z(2)$ for 
$N = 4k$, and $\Z(4)$ for $N = 4k + 2$. The center of the group $Sp(N)$ is
$\Z(2)$ for all $N$. Hence, the universal covering groups of all main sequence 
Lie groups have a non-trivial center. What about the exceptional groups? 
Interestingly, the groups $G(2)$, $F(4)$, and $E(8)$ have a trivial center and
are their own universal covering groups. The groups $E(6)$ and $E(7)$, on the 
other hand, have the non-trivial centers $\Z(3)$ and $\Z(2)$, respectively. The
exceptional Lie group $G(2)$ is the simplest group whose universal covering 
group has a trivial center.

The triviality of the center has profound consequences for the way in which
confinement is realized. In particular, a static ``quark'' in the fundamental 
$\{7\}$ representation of $G(2)$ can be screened by three $G(2)$ ``gluons'' in 
the adjoint $\{14\}$ representation. As a result, in $G(2)$ Yang-Mills theory 
the color flux string connecting two static $G(2)$ ``quarks'' can break due to 
the creation of dynamical gluons. This phenomenon is reminiscent of full QCD 
(with an $SU(3)$ color gauge group) in which the string connecting a static 
quark and anti-quark can break due to the pair creation of light dynamical 
quarks. Indeed, 6 of the 14 $G(2)$ gluons transform as $\{3\}$ and 
$\{\overline 3\}$ under the $SU(3)$ subgroup of $G(2)$ and thus qualitatively 
behave like dynamical quarks and anti-quarks. In particular, they explicitly 
break the $\Z(3)$ center symmetry of the $SU(3)$ subgroup down to the trivial 
center of $G(2)$. The remaining $14 - 6 = 8$ $G(2)$ ``gluons'' transform as 
$\{8\}$ under the $SU(3)$ subgroup and hence resemble the ordinary gluons 
familiar from QCD. It should be pointed out that --- despite the broken string
--- just like full QCD, $G(2)$ Yang-Mills theory is still expected to confine 
color. In particular, one does not expect colored states of single $G(2)$ 
``gluons'' in the physical spectrum. The triviality of the center of $G(2)$ 
Yang-Mills theory also affects the physics at high temperatures. In particular,
there is not necessarily a deconfinement phase transition, and we expect merely
a crossover between a low-temperature ``glueball'' regime and a 
high-temperature $G(2)$ ``gluon'' plasma. Due to the triviality of the center, 
unlike e.g. for $SU(N_c)$ Yang-Mills theory, there is no qualitative difference
between the low- and the high-temperature regimes because the Polyakov loop is 
no longer a good order parameter.

It is often being asked which degrees of freedom are responsible for 
confinement. Popular candidates are dense instantons, merons, Abelian monopoles
and center vortices. Center vortices (and 't Hooft twist sectors) are absent in
$G(2)$ gauge theories, while
the other topological objects potentially exist, although their identification 
is a very subtle issue that often involves unsatisfactory gauge fixing 
procedures. At strong coupling $G(2)$ lattice gauge theories still confine 
without a center. Hence, center vortices should not be necessary to explain
the absence of colored states in the physical spectrum \cite{Rei01}. Still, the
center plays an important role for the finite temperature deconfinement phase 
transition in $SU(N_c)$ Yang-Mills theory, and center vortices may well be 
relevant for this physics. If $G(2)$ Yang-Mills theory indeed has no finite 
temperature deconfinement phase transition, one might argue that this is due to
the absence of center vortices and twist sectors. Assuming that they can be 
properly defined, 
Abelian monopoles are potentially present in $G(2)$ gauge theory, and might be
responsible for the absence of colored states. On the other hand, if ---
despite of the existence of Abelian monopoles --- a deconfinement phase 
transition does not exist in $G(2)$ Yang-Mills theory, monopoles might not be 
responsible for the physics of deconfinement. In any case, quantifying these 
issues in a concrete way is a very difficult task.

The exceptional confinement in $G(2)$ gauge theory can be smoothly connected 
with the usual $SU(3)$ confinement by exploiting the Higgs mechanism. When a 
scalar field in the fundamental $\{7\}$ representation of $G(2)$ picks up a
vacuum expectation value, the gauge symmetry breaks down to $SU(3)$, and the
6 additional $G(2)$ ``gluons'' become massive. By progressively increasing the 
vacuum expectation value of the Higgs field, one can decouple those particles, 
thus smoothly interpolating between $G(2)$ and $SU(3)$ gauge theories. In this 
way, we use $G(2)$ gauge theories as a theoretical laboratory in which the 
$SU(3)$ theories we are most interested in are embedded in an unusual 
environment. This provides theoretical insight not only into the exceptional 
$G(2)$ confinement, but also into the $SU(3)$ confinement that occurs in 
Nature.

The rest of the paper is organized as follows. In section 2 we review the 
center symmetry, the construction of the Polyakov loop, and some subtle issues 
related to the physics of non-Abelian gauge fields in a finite volume. Some
details of periodic and $C$-periodic boundary conditions are discussed in an
appendix. In section 3 we present the basic features of the exceptional group 
$G(2)$. Section 4 contains the discussion of various field theories with gauge 
group $G(2)$. As a starting point, we consider $G(2)$ Yang-Mills theory, which 
we then break to the $SU(3)$ subgroup using the Higgs mechanism. We then add 
fermion fields in both the fundamental and the adjoint representation, thus 
arriving at $G(2)$ QCD and supersymmetric $G(2)$ Yang-Mills theory, 
respectively. In both cases, we concentrate on the chiral symmetries and we 
discuss how they are realized at low and at high temperature. In section 5 we 
substantiate the qualitative pictures painted in section 4 by performing strong
coupling calculations in $G(2)$ lattice gauge theory. In particular, we show 
that the theory confines although there is no string tension. Finally, section 
6 contains our conclusions.

\section{Center Symmetry, Polyakov Loop and Gauge Fields in a Finite Volume}

When defined properly, the Polyakov loop is a useful order parameter in
Yang-Mills gauge theories with a non-trivial center symmetry, which 
distinguishes confinement at low temperatures from deconfinement at high 
temperatures. In particular, the Polyakov loop varies under non-trivial 
transformations in the center of the gauge group and it thus signals the 
spontaneous breakdown of the center symmetry at high temperatures. The 
expectation value of the Polyakov loop $\langle \Phi \rangle = \exp(- \beta F)$
measures the free energy $F$ of an external static test quark. In a confined 
phase with unbroken non-trivial center symmetry the free energy of a static 
quark is infinite. Hence, $\langle \Phi \rangle = 0$ and the center symmetry is
unbroken. In a deconfined phase, on the other hand, $F$ is finite, 
$\langle \Phi \rangle \neq 0$, and the center symmetry is spontaneously broken.

The Polyakov loop is a rather subtle observable whose definition needs special
care. In particular, it is sensitive to spatial and temporal boundary 
conditions. For example, for a system of $SU(3)$ Yang-Mills gluons on a finite
torus, the expectation value of the Polyakov loop is always zero even in the 
deconfined phase \cite{Hil83}. This is a consequence of the $\Z(3)$ Gauss law: 
a single static test quark cannot exist in a periodic volume because its color 
flux cannot go to infinity and must thus end in an anti-quark. Due to the Gauss
law, a torus is always neutral. Since it always vanishes, on a finite torus the
expectation value of the Polyakov loop does not contain any useful information 
about confinement or deconfinement. Still, using the Polyakov loop, one can, 
for example, define its finite volume constraint effective potential, which 
does indeed allow one to distinguish confined from deconfined phases.

Let us consider a non-Abelian Yang-Mills theory with gauge group $G$ and
anti-Hermitean vector potential $A_\mu(x)$. The physics is invariant under 
non-Abelian gauge transformations
\begin{equation}
A_\mu(x)' = \Omega(x) (A_\mu(x) + \p_\mu) \Omega(x)^\dagger,
\end{equation}
where $\Omega(x) \in G$. We now put the system in a finite 4-dimensional
rectangular space-time volume of size $L_1 \times L_2 \times L_3 \times L_4$. 
Here $L_i$ is the extent in the spatial $i$-direction and $L_4 = \beta = 1/T$
is the extent of periodic Euclidean time which determines the inverse 
temperature $\beta = 1/T$. We consider periodic boundary conditions in both 
space and Euclidean time, such that our 4-dimensional space-time volume is a 
hyper-torus. This means that gauge-invariant physical quantities --- but not 
the gauge-dependent vector potentials themselves --- are periodic functions of 
space-time. The gauge fields themselves must be periodic only up to gauge 
transformations, i.e.
\begin{equation}
A_\mu(x + L_\nu e_\nu) = 
\Omega_\nu(x) (A_\mu(x) + \p_\mu) \Omega_\nu(x)^\dagger.
\end{equation}
Here $e_\nu$ is the unit-vector in the $\nu$-direction and $\Omega_\nu(x)$ is
a gauge transformation that relates the gauge field $A_\mu(x + L_\nu e_\nu)$,
shifted by a distance $L_\nu$ in the $\nu$-direction, to the unshifted gauge
field $A_\mu(x)$. Mathematically speaking, the $\Omega_\nu(x)$ define a 
universal fiber bundle of transition functions which glue the torus together at
the boundaries. As explained in the appendix, the transition functions must
obey the cocycle condition
\begin{equation}
\label{cocycle}
\Omega_\nu(x + L_\rho e_\rho) \Omega_\rho(x) = 
\Omega_\rho(x + L_\nu e_\nu) \Omega_\nu(x) z_{\nu\rho}.
\end{equation}
This consistency conditions contains the twist-tensor $z_{\nu\rho}$ which takes
values in the center of the gauge group.

It should be noted that the transition functions $\Omega_\nu(x)$ are physical 
degrees of freedom of the non-Abelian gauge field, just like the vector
potentials $A_\mu(x)$ themselves. In particular, in the path integral one must
also integrate over the transition functions, otherwise gauge-variant
unphysical quantities like $A_\mu(x)$ itself might also become periodic. Under 
general (not necessarily periodic) gauge transformations $\Omega(x)$ the 
transition functions transform as
\begin{equation}
\label{transition}
\Omega_\nu(x)' = \Omega(x + L_\nu e_\nu) \Omega_\nu(x) \Omega(x)^\dagger.
\end{equation}
In lattice gauge theory the transition functions are nothing but the Wilson
parallel transporters on the links that connect two opposite sides of the 
periodic volume. 

The twist-tensor is gauge-invariant. Hence, as was first pointed out by 
't Hooft \cite{tHo77}, non-Abelian gauge fields on a torus fall into gauge 
equivalence classes characterized by the twist-tensor, which provides a 
gauge-invariant characterization of distinct superselection sectors. A 
non-trivial twist-tensor $z_{\nu\rho} \neq 0$ implies background electric or 
magnetic fluxes that wrap around the torus in various directions, while the 
sector with trivial twist $z_{\nu\rho} = 1$ describes a periodic world without 
electric or magnetic fluxes. It should be noted that one need not sum over the 
different twist-sectors in the path integral, because they correspond to 
distinct superselection sectors of the theory.

Interestingly, in a non-Abelian Yang-Mills theory there is a symmetry
transformation
\begin{equation}
\label{centersym}
\Omega_\mu(x)' = \Omega_\mu(x) z_\mu,
\end{equation}
which leaves the boundary condition as well as the twist-tensor --- and hence 
the superselection sector --- invariant. Here $z_\mu$ is an element of the 
center of the gauge group $G$. This center symmetry exists only if all fields 
in the theory are center-blind. This is automatically the case for the gauge 
fields which transform in the adjoint representation. However, if there are 
fields that transform non-trivially under the center, the center symmetry is 
explicitly broken. For example, a matter field that transforms as
\begin{equation}
\Psi(x)' = \Omega(x) \Psi(x),
\end{equation}
under gauge transformations, obeys the boundary condition
\begin{equation}
\Psi(x + L_\mu e_\mu) = \Omega_\mu(x) \Psi(x),
\end{equation}
which is gauge-covariant, but not invariant under the center symmetry of 
eq.(\ref{centersym}).

Based on the previous discussion, we are finally ready to attempt a first
definition of the Polyakov loop
\begin{equation}
\label{Polyakov}
\Phi(\vec x) = \mbox{Tr}[\Omega_4(\vec x,0)
{\cal P} \exp(\int_0^\beta dt \ A_4(\vec x,t))],
\end{equation}
which is invariant under the transformations of eq.(\ref{transition}) only 
because the transition function $\Omega_4(\vec x,0)$ is included in its 
definition. Then one obtains
\begin{eqnarray}
\Phi(\vec x)'&=&\mbox{Tr}[\Omega_4(\vec x,0)'
{\cal P} \exp(\int_0^\beta dt \ A_4(\vec x,t)')] 
\nonumber \\
&=&\mbox{Tr}[\Omega(\vec x,\beta) \Omega_4(\vec x,0) \Omega(\vec x,0)^\dagger
\Omega(\vec x,0) {\cal P} \exp(\int_0^\beta dt \ A_4(\vec x,t))
\Omega(\vec x,\beta)^\dagger ] \nonumber \\
&=&\Phi(\vec x).
\end{eqnarray}
Under the center symmetry transformation of eq.(\ref{centersym}) the Polyakov 
loop transforms as
\begin{equation}
\Phi(\vec x)' = \Phi(\vec x) z_4,
\end{equation}
and thus it provides us with an order parameter for the spontaneous breakdown 
of the center symmetry. However, on a torus the expectation value of the 
Polyakov loop $\langle \Phi \rangle$ always vanishes, simply because 
spontaneous symmetry breaking --- in the sense of a non-vanishing order 
parameter --- does not occur in a finite volume. Alternatively, one may say 
that the expectation value of the Polyakov loop always vanishes because the 
presence of a single static quark is incompatible with the Gauss law on a 
torus. In any case, since it is always zero, on a finite torus the expectation 
value of the Polyakov loop does not contain any information about confinement 
or deconfinement, or about how the center symmetry is dynamically realized. 

Still, even on a finite torus the Polyakov loop can be used to define related 
quantities that indeed contain useful information about confinement versus
deconfinement, and about the realization of the center symmetry. For example,
one can define the finite volume ($\beta V = L_1 L_2 L_3 L_4$) constraint 
effective potential ${\cal V}(\Phi)$ of the Polyakov loop as
\begin{eqnarray}
\exp(- \beta V {\cal V}(\Phi))&=&\int {\cal D}A \ 
\delta(\Phi - \frac{1}{V} \int d^3x \ \mbox{Tr}[\Omega_4(\vec x,0)
{\cal P} \exp(\int_0^\beta dt \ A_4(\vec x,t))]) \nonumber \\
&\times& \exp(-S[A]),
\end{eqnarray}
where $S[A]$ is the Euclidean Yang-Mills action. In the confined phase, the 
constraint effective potential ${\cal V}(\Phi)$ has its minimum at $\Phi = 0$,
while in the deconfined phase it has degenerate minima at $\Phi \neq 0$ which
are related to one another by center symmetry transformations.

One may still not be satisfied with the previous definition of the Polyakov
loop. In particular, one may argue that the center symmetry transformations are
part of the gauge group. In that case, the Polyakov loop, as defined in 
eq.(\ref{Polyakov}), would simply be a gauge-variant unphysical quantity. In
order to resurrect the Polyakov loop from this deadly argument, we now discuss
a space-time volume with $C$-periodic boundary conditions in the spatial 
directions. Thermodynamics dictates that the boundary conditions in the
Euclidean time direction remain periodic. Even the expectation value of the 
Polyakov loop itself becomes a useful observable when $C$-periodic boundary 
conditions are used in $SU(3)$ Yang-Mills theory. In that case, a spatial shift
by a distance $L_i$ is accompanied by a charge-conjugation transformation 
\cite{Pol91,Kro91}. A single static quark can exist in a $C$-periodic box 
because its color flux can end in a mirror anti-quark on the other side of the 
boundary. As a consequence, the expectation value of the Polyakov loop no 
longer vanishes automatically \cite{Wie92}. In fact, it now vanishes only if 
one takes the infinite volume limit in the confined phase, while it remains 
non-zero in the deconfined phase. 

In a $C$-periodic volume the physics is periodic up to a charge-conjugation
twist, i.e. all physical quantities are replaced by their charge-conjugates
when shifted by a distance $L_i$ in a spatial direction. Of course, the gauge 
fields themselves are $C$-periodic only up to gauge transformations, i.e.
\begin{equation}
A_\mu(x + L_i e_i) = 
\Omega_i(x) (A_\mu(x)^* + \p_\mu) \Omega_i(x)^\dagger.
\end{equation}
Here $A_\mu(x)^*$ is the charge-conjugate of the gauge field $A_\mu(x)$. In the
Euclidean time direction we keep periodic boundary conditions, i.e.
\begin{equation}
A_\mu(x + \beta e_i) = 
\Omega_4(x) (A_\mu(x) + \p_\mu) \Omega_4(x)^\dagger.
\end{equation}

As shown in the appendix, the cocycle conditions for $C$-periodic boundary 
conditions are given by
\begin{eqnarray}
\label{Ccocycle}
&&\Omega_i(x + L_j e_j) \Omega_j(x)^* = 
\Omega_j(x + L_i e_i) \Omega_i(x)^* z_{ij}, \nonumber \\
&&\Omega_i(x + \beta e_4) \Omega_4(x)^* = 
\Omega_4(x + L_i e_i) \Omega_i(x) z_{i4},
\end{eqnarray}
and hence they differ from those for periodic boundary conditions. With 
$C$-periodic boundary conditions the transition functions transform under gauge
transformations as
\begin{equation}
\label{Ctransition}
\Omega_i(x)' = \Omega(x + L_i e_i) \Omega_i(x) \Omega(x)^T, \
\Omega_4(x)' = \Omega(x + \beta e_4) \Omega_4(x) \Omega(x)^\dagger,
\end{equation}
where $T$ denotes the transpose. As we work out in the appendix, unlike for 
periodic boundary conditions, there are constraints on the twist-tensor itself.
First
\begin{equation}
\label{Cmtwist}
z_{ij}^2 z_{jk}^2 z_{ki}^2 = 1,
\end{equation}
and second
\begin{equation}
\label{Cetwist}
z_{i4}^2 = z_{j4}^2.
\end{equation}
Interestingly, with $C$-periodic boundary conditions the twist-tensor is no
longer invariant against the center symmetry transformations of 
eq.(\ref{centersym}). One finds
\begin{equation}
\label{Cgtwist}
z_{ij}' = z_{ij} z_i^2 {z_j^*}^2, \ 
z_{i4}' = z_{i4} {z_4^*}^2.
\end{equation} 
These relations can be used to relate gauge-equivalent twist-tensors to one
another. 

First, let us assume that the center of the gauge group $G$ is $\Z(N_c)$ with 
odd $N_c$. This is the case for $SU(N_c)$ groups with odd $N_c$ as well as for 
$E(6)$ which has the center $\Z(3)$. Of course, the physical color gauge group 
$SU(3)$ with its center $\Z(3)$ also falls in this class. In that case, in a 
$C$-periodic volume all twist-sectors are gauge-equivalent. In particular, 
using eq.(\ref{Cetwist}) and putting the transformation parameter to 
$z_4^2 = z_{i4}$ one obtains $z_{i4}' = 1$ for all $i$. Next, we put the 
transformation parameter $z_1 = 1$ and we choose $z_2^2 = z_{12}$ such that 
$z_{12}' = 1$, and $z_3^2 = z_{13}$ such that $z_{13}' = 1$. Using the 
consistency condition eq.(\ref{Cmtwist}) one finally obtains 
${z_{23}'}^2 = {z_{21}'}^2 {z_{13}'}^2 = 1$ such that $z_{23}' = 1$. Hence, in 
this case the entire twist-tensor $z_{\mu\nu}' = 1$ is trivial. Consequently, 
for gauge groups with the center $\Z(N_c)$ with odd $N_c$ there is only one 
$C$-periodic boundary condition. In other words, for $C$-periodic boundary
conditions no analog of 't Hooft's electric and magnetic flux sectors exists 
--- all twist-sectors are gauge-equivalent to the trivial one. In these cases, 
according to eq.(\ref{Cetwist}) the twist-tensor element $z_{i4}$ is 
independent of the spatial direction $i$. Furthermore, since $N_c$ is odd, its
square-root in the center $\sqrt{z_{i4}} \in \Z(N_c)$ is uniquely defined 
(without any sign-ambiguity). According to eq.(\ref{Cgtwist}) it transforms as
\begin{equation}
\label{Cgsqrt}
\sqrt{z_{i4}'} = \sqrt{z_{i4}} z_4^*,
\end{equation}
under center transformations. This finally allows us to write down a 
completely gauge-invariant definition of the Polyakov loop in a $C$-periodic 
volume
\begin{equation}
\Phi(\vec x) = \mbox{Tr}[\sqrt{z_{i4}} \Omega_4(\vec x,0)
{\cal P} \exp(\int_0^\beta dt \ A_4(\vec x,t))].
\end{equation}
Unlike for periodic boundary conditions, the center transformation of the
transition function $\Omega_4(\vec x,0)' = \Omega_4(\vec x,0) z_4$ is now 
compensated by the variation of the square-root of the twist-tensor from
eq.(\ref{Cgsqrt}) and one obtains $\Phi(\vec x)' = \Phi(\vec x)$.
Defined in this completely gauge-invariant way, the expectation value of the 
Polyakov loop $\langle \Phi \rangle = \exp(- \beta F)$ indeed determines the 
free energy $F$ of a single static quark. In contrast to the periodic torus, a
$C$-periodic volume can contain a single static quark, because the color flux
string emanating from it can end in a charge-conjugate anti-quark on the other 
side of the boundary. 

The groups $SU(N_c)$ with even $N_c$ have a sign-ambiguity in the definition of
the square-root of a center element. In that case, the expectation value of the
Polyakov loop vanishes even in a $C$-periodic volume. The same is true for 
$Spin(N)$ --- the universal covering group of $SO(N)$ --- which has the center
$\Z(2)$ for odd $N$, $\Z(2) \otimes \Z(2)$ for $N = 4 k$, and $\Z(4)$ for 
$N = 4 k + 2$, as well as for the symplectic groups $Sp(N)$ and the exceptional
group $E(7)$ which both have the center $\Z(2)$. In those cases, one is limited
to the construction of the finite volume constraint effective potential. The 
exceptional groups $G(2)$, $F(4)$ and $E(8)$ have a trivial center. Then the
Polyakov loop is not an order parameter, but it can at least be defined without
any problems even in a simple periodic volume.

Keeping in mind the subtleties discussed above, when we refer to the Polyakov 
loop in the rest of this paper, strictly speaking, we mean the location of a 
minimum of its constraint effective potential on the torus, or its expectation 
value in a $C$-periodic box. Both are identical in the infinite volume limit.

\section{The Exceptional Group $G(2)$}

In this section we discuss some basic properties of the Lie group $G(2)$ ---
the simplest among the exceptional groups $G(2)$, $F(4)$, $E(6)$, $E(7)$ and
$E(8)$ --- which do not fit into the main sequences $SO(N) \simeq Spin(N)$, 
$SU(N)$ and $Sp(N)$. While there is only one non-Abelian compact Lie algebra of
rank 1 --- namely the one of $SO(3) \simeq SU(2) = Sp(1)$ --- there are four of
rank 2. These rank 2 algebras generate the groups $G(2)$, $SO(5) \simeq Sp(2)$,
$SU(3)$ and $SO(4) \simeq SU(2) \otimes SU(2)$, which have 14, 10, 8 and 6 
generators, respectively. For us the group $G(2)$ is of particular interest 
because it has a trivial center and is its own universal covering group. As we 
will see later, this has interesting consequences for the confinement 
mechanism.

It is natural to construct $G(2)$ as a subgroup of $SO(7)$ which has rank 3 and
21 generators. The $7 \times 7$ real orthogonal matrices $\Omega$ of the group 
$SO(7)$ have determinant 1 and obey the constraint
\begin{equation}
\Omega_{ab} \Omega_{ac} = \delta_{bc}.
\end{equation}
The $G(2)$ subgroup contains those matrices that, in addition, satisfy the 
cubic constraint
\begin{equation}
\label{cubic}
T_{abc} = T_{def} \Omega_{da} \Omega_{eb} \Omega_{fc}.
\end{equation}
Here $T$ is a totally anti-symmetric tensor whose non-zero elements follow by 
anti-symmetrization from
\begin{equation}
\label{tensor}
T_{127} = T_{154} = T_{163} = T_{235} = T_{264} = T_{374} = T_{576} = 1.
\end{equation}
The tensor $T$ also defines the multiplication rules for octonions 
\cite{Gun73}. 
Eq.(\ref{tensor}) implies that eq.(\ref{cubic}) represents 7 non-trivial
constraints which reduce the 21 degrees of freedom of $SO(7)$ to the 14 
parameters of $G(2)$. It should be noted that $G(2)$ inherits the reality
properties of $SO(7)$: all its representations are real.

We make the following choice for the first 8 generators of $G(2)$ in the 
7-di\-men\-si\-o\-nal fundamental representation \cite{Gun73}
\begin{equation}
\label{su3gen}
\Lambda_a = \frac{1}{\sqrt{2}} \left( \begin{array}{ccc} \lambda_a & 0 & 0 \\ 
0 & \; -\lambda_a^* & 0\\ 0 & 0 & 0 \
\end{array} \right).
\end{equation}
Here $\lambda_a$ (with $a \in \{1,2,...,8\}$) are the usual $3\times 3$ 
Gell-Mann generators of $SU(3)$ which indeed is a subgroup of $G(2)$. We have
chosen the standard normalization $\mbox{Tr} \lambda_a \lambda_b = 
\mbox{Tr} \Lambda_a \Lambda_b = 2 \delta_{ab}$. The representation we have 
chosen involves complex numbers. However, it is unitarily equivalent to a 
representation that is entirely real. In the chosen basis of the generators it 
is manifest that under $SU(3)$ subgroup transformations the 7-dimensional 
representation decomposes into
\begin{equation}
\label{dec7}
\{7\} = \{3\} \oplus \{\overline 3\} \oplus \{1\} .
\end{equation}
Since $G(2)$ has rank 2, only two generators can be diagonalized 
simultaneously. In our choice of basis these are the $SU(3)$ subgroup 
generators $\Lambda_3$ and $\Lambda_8$. Consequently, just as for $SU(3)$, the 
weight diagrams of $G(2)$ representations can be drawn in a 2-dimensional 
plane. For example, the weight diagram of the fundamental representation is 
shown in figure 1. 
\begin{figure}[htb]
\begin{center}
\epsfig{file=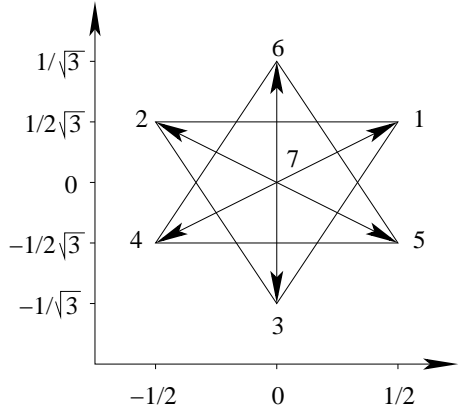,width=60mm,angle=0}
\caption{\it The weight diagram of the 7-dimensional fundamental representation
of $G(2)$ (rescaled by a factor $\sqrt{2}$).}
\end{center}    
\end{figure}
One notes that it is indeed a superposition of the weight diagrams of a 
$\{3\}$, $\{\overline 3 \}$ and $\{1\}$ in $SU(3)$. Since all $G(2)$ 
representation are real, the $\{7\}$ representation is equivalent to its 
complex conjugate. As a consequence, $G(2)$ ``quarks'' and ``anti-quarks'' are 
indistinguishable. In particular, a $G(2)$ ``quark'' $\{7\}$ consists of an 
$SU(3)$ quark $\{3\}$, an $SU(3)$ anti-quark $\{\overline 3\}$ and an $SU(3)$ 
singlet $\{1\}$. It should be noted that the $\{3\} \oplus \{\overline 3 \}$ 
contained in the $\{7\}$ of $G(2)$ corresponds to a real reducible 
6-dimensional representation of $SU(3)$.

As usual,
\begin{eqnarray}
T^+&=&\frac{1}{\sqrt{2}}(\Lambda_1 + i \Lambda_2) = \kbra{1}{2} - \kbra{5}{4},
\nonumber \\  
T^-&=&\frac{1}{\sqrt{2}}(\Lambda_1 - i \Lambda_2) = \kbra{2}{1} - \kbra{4}{5},
\nonumber \\
U^+&=&\frac{1}{\sqrt{2}}(\Lambda_4 + i \Lambda_5) = \kbra{2}{3} - \kbra{6}{5},
\nonumber \\
U^-&=&\frac{1}{\sqrt{2}}(\Lambda_4 - i \Lambda_5) = \kbra{3}{2} - \kbra{5}{6},
\nonumber \\
V^+&=&\frac{1}{\sqrt{2}}(\Lambda_4 + i \Lambda_6) = \kbra{1}{3} - \kbra{6}{4},
\nonumber \\
V^-&=&\frac{1}{\sqrt{2}}(\Lambda_6 - i \Lambda_4) = \kbra{3}{1} - \kbra{4}{6},
\end{eqnarray}
define $SU(3)$ shift operations between the different states $|1\rangle$,
$|2\rangle$,...,$|7\rangle$ in the fundamental representation. The remaining 6 
generators of $G(2)$ also define shifts
\begin{eqnarray}
\label{g2gen}
X^+&=&\frac{1}{\sqrt{2}}(\Lambda_9 + i \Lambda_{10})    =
\kbra{2}{4} -\kbra{1}{5} -\sqrt{2}\kbra{7}{3} -\sqrt{2}\kbra{6}{7} , 
\nonumber \\
X^-&=&\frac{1}{\sqrt{2}}(\Lambda_9 - i \Lambda_{10})    =
\kbra{4}{2} -\kbra{5}{1} -\sqrt{2}\kbra{3}{7} -\sqrt{2}\kbra{7}{6} , 
\nonumber \\
Y^+&=&\frac{1}{\sqrt{2}}(\Lambda_{11} + i \Lambda_{12}) = 
\kbra{6}{1} - \kbra{4}{3} -\sqrt{2}\kbra{2}{7} -\sqrt{2}\kbra{7}{5} , 
\nonumber \\
Y^-&=&\frac{1}{\sqrt{2}}(\Lambda_{11} - i \Lambda_{12}) = 
\kbra{1}{6} - \kbra{3}{4} -\sqrt{2}\kbra{7}{2} -\sqrt{2}\kbra{5}{7} , 
\nonumber \\
Z^+&=&\frac{1}{\sqrt{2}}(\Lambda_{13} + i \Lambda_{14}) = 
\kbra{3}{5} - \kbra{2}{6} -\sqrt{2}\kbra{7}{1} -\sqrt{2}\kbra{4}{7} , 
\nonumber \\
Z^-&=&\frac{1}{\sqrt{2}}(\Lambda_{13} - i \Lambda_{14}) = 
\kbra{5}{3} - \kbra{6}{2} -\sqrt{2}\kbra{1}{7} -\sqrt{2}\kbra{7}{4}.
\end{eqnarray}
The generators themselves transform under the 14-dimensional adjoint
representation of $G(2)$ whose weight diagram is shown in figure 2. 
\begin{figure}[htb]
\begin{center}
\epsfig{file=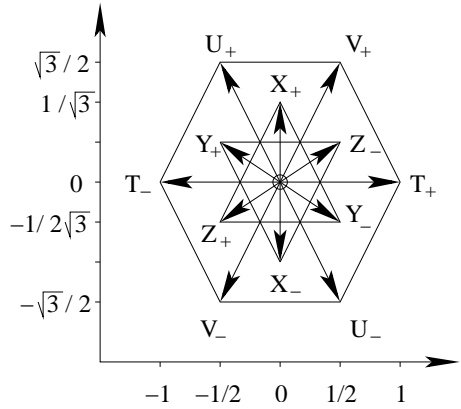,width=60mm,angle=0}
\caption{\it The weight diagram of the 14-dimensional adjoint representation of
$G(2)$ (rescaled by a factor $\sqrt{2}$).}
\end{center}    
\end{figure}
From this diagram one sees that under an $SU(3)$ subgroup transformation the 
adjoint representation of $G(2)$ decomposes into
\begin{equation}
\label{dec14}
\{14\} = \{8\} \oplus \{3\} \oplus \{\overline 3\}.
\end{equation}
This implies that $G(2)$ ``gluons'' $\{14\}$ consist of the usual $SU(3)$ 
gluons $\{8\}$ as well as of 6 additional ``gluons'' with $SU(3)$ quark and 
anti-quark color quantum numbers $\{3\}$ and $\{\overline 3\}$.

Let us now discuss the center of $G(2)$. It is interesting to note that the
maximal Abelian (Cartan) subgroup of both $G(2)$ and $SU(3)$ is $U(1)^2$ which
must contain the center in both cases. Since $G(2)$ contains $SU(3)$ as a
subgroup its center cannot be bigger than $\Z(3)$ (the center of $SU(3)$)
because the potential center elements of $G(2)$ must commute with all $G(2)$
matrices (not just with the elements of the $SU(3)$ subgroup). In the 
fundamental representation of $G(2)$ the center elements of the $SU(3)$ 
subgroup are given by
\begin{equation}
Z = \left( \begin{array}{ccc} z \1 & 0 & 0 \\ 
0 & \; z^*\1 & 0 \\0 & 0 & \; 1 \ \end{array} \right).
\end{equation}
where $\1$ is the $3 \times 3$ unit matrix and $z \in \{1,\exp(\pm 2 \pi i/3\}$
is an element of $\Z(3)$. By construction, the three $7 \times 7$ matrices $Z$ 
commute with the 8 generators of the $SU(3)$ subgroup of $G(2)$. However, an 
explicit calculation shows that this is not the case for the remaining 6 
generators. Consequently, the center of $G(2)$ is trivial and contains only the
identity. The above argument applies to any representation of $G(2)$. In other 
words, the universal covering group of $G(2)$ is $G(2)$ itself and still it has
a trivial center. As we will see, this has drastic consequences for 
confinement. In particular, the string between static $G(2)$ ``quarks'' can 
break already in the pure gauge theory through the creation of dynamical 
``gluons''.

In $SU(3)$ the non-trivial center $\Z(3)$ gives rise to the concept of 
triality. For example, the trivial representation $\{1\}$ and the adjoint
representation $\{8\}$ of $SU(3)$ have trivial triality, while the fundamental
$\{3\}$ and anti-fundamental $\{\overline 3\}$ have non-trivial opposite
trialities. Since its center is trivial, the concept of triality does not
extend to $G(2)$. In particular, as one can see from
eqs.(\ref{dec7},\ref{dec14}), $G(2)$ representations decompose into mixtures of
$SU(3)$ representations with different trialities. This has interesting 
consequences for the results of $G(2)$ tensor decompositions \cite{McKay}.
For example, in contrast to the $SU(3)$ case, the product of two fundamental 
representations
\begin{equation}
\label{prod77}
\{7\} \otimes \{7\} = \{1\} \oplus \{7\} \oplus \{14\} \oplus \{27\}, 
\end{equation}
contains both the trivial and the adjoint representation. The $\{1\}$ and
$\{27\}$ representations are symmetric under the exchange of the two $\{7\}$
representations, while $\{7\}$ and $\{14\}$ are anti-symmetric. As a result of
eq.(\ref{prod77}), already two $G(2)$ ``quarks'' can form a color-singlet. Just
as for $SU(3)$, three $G(2)$ ``quarks'' can form a color-singlet ``baryon''
because
\begin{equation}
\label{prod}
\{7\} \otimes \{7\} \otimes \{7\} = \{1\} \oplus 4 \; \{7\} \oplus 
2 \; \{14\} \oplus 3 \; \{27\} \oplus 2 \; \{64\} \oplus 3 \; \{77\}.
\end{equation}
Another interesting example is
\begin{eqnarray}
\label{product}
\{14\} \otimes \{14\} \otimes \{14\}&=&\{1\} \oplus \{7\} \oplus 5 \; \{14\} 
\oplus 3 \; \{27\} \oplus 2 \; \{64\} \oplus 4 \; \{77\} \oplus 3 \; \{77'\}
\nonumber \\
&\oplus&\{182\} \oplus 3 \; \{189\} \oplus \{273\} \oplus 2 \; \{448\}.
\end{eqnarray}
As a consequence of the absence of triality, the decomposition of the tensor 
product of three adjoint representations contains the fundamental
representation. This means that three $G(2)$ ``gluons'' $G$ can screen a single
$G(2)$ ``quark'' $q$, and thus a color-singlet hybrid $qGGG$ can be formed. 
Later we will also need the results for further tensor product decompositions, 
two of which are listed here
\begin{eqnarray}
&&\{7\} \otimes \{14\} = \{7\} \oplus \{27\} \oplus \{64\}, 
\nonumber \\
&&\{14\} \otimes \{14\} = \{1\} \oplus \{14\} \oplus \{27\} \oplus 
\{77\} \oplus \{77'\}.
\end{eqnarray}

It is also interesting to consider the homotopy groups related to $G(2)$
because this tells us what kind of topological excitations can arise. As for
$SU(3)$, the third homotopy group of $G(2)$ is
\begin{equation}
\Pi_3[G(2)] = \Z.
\end{equation}
Hence, there are $G(2)$ instantons of any additive integer topological charge
and, consequently, also a $\theta$-vacuum angle. Another homotopy group of
interest is
\begin{equation}
\Pi_2[G(2)/U(1)^2] = \Pi_1[U(1)^2] = \Z^2.
\end{equation}
Again, this is just like for $SU(3)$. Physically, this means that 
't Hooft-Polyakov monopoles with two kinds of magnetic charge show up when 
$G(2)$ is broken to its maximal Abelian (Cartan) subgroup $U(1)^2$. For $SU(3)$
with center $\Z(3)$ the homotopy
\begin{equation}
\Pi_1[SU(3)/\Z(3)] = \Pi_0[\Z(3)] = \Z(3),
\end{equation}
implies that the pure gauge theory has non-trivial twist-sectors. 
Interestingly, in contrast to $SU(3)$, for $G(2)$ which has a trivial center 
$I = \{\1\}$ the first homotopy group
\begin{equation}
\Pi_1[G(2)/I] = \Pi_0[I] = \{0\}
\end{equation}
is trivial. Hence, even in the pure gauge theory non-trivial twist-sectors do
not exit.

\section{$G(2)$ Gauge Theories}

In this section we discuss various theories with gauge group $G(2)$. We start 
with pure Yang-Mills theory and then add charged matter fields in various
representations. For example, we consider a scalar Higgs field in the 
fundamental representation which breaks $G(2)$ down to $SU(3)$. By varying 
the vacuum expectation value of the Higgs field one can interpolate between a 
$G(2)$ and an $SU(3)$ gauge theory. We also add Majorana ``quarks'' first in 
the fundamental $\{7\}$ representation and then also in the adjoint $\{14\}$ 
representation. The former theory is closely related to $SU(3)$ QCD, while the 
latter corresponds to ${\cal N} = 1$ supersymmetric $G(2)$ Yang-Mills theory. 

\subsection{$G(2)$ Yang-Mills Theory}

Let us first consider the simplest $G(2)$ gauge theory --- $G(2)$ Yang-Mills
theory. Since $G(2)$ has 14 generators there are 14 ``gluons''. Under the
subgroup $SU(3)$ 8 of them transform as ordinary gluons, i.e. as an $\{8\}$
of $SU(3)$. The remaining 6 $G(2)$ gauge bosons break-up into $\{3\}$ and
$\{\overline 3 \}$, i.e. they have the color quantum numbers of ordinary quarks
and anti-quarks. Of course, in contrast to real quarks, these objects are 
bosons with spin 1. Still, these additional 6 ``gluons'' have somewhat similar
effects as quarks in full QCD. In particular, they explicitly break the $\Z(3)$
center symmetry of $SU(3)$ and make the center symmetry of $G(2)$ Yang-Mills 
theory trivial. The Lagrangian for $G(2)$ Yang-Mills theory takes the standard
form
\begin{equation}
{\cal L}_{YM}[A] = \frac{1}{2 g^2} \mbox{Tr} F_{\mu\nu} F_{\mu\nu},
\end{equation}
where the field strength
\begin{equation}
F_{\mu\nu} = \p_\mu A_\nu - \p_\nu A_\mu + [A_\mu,A_\nu],
\end{equation}
is derived from the vector potential
\begin{equation}
A_\mu(x) = i g A_\mu^a(x) \Lambda_a.
\end{equation}
The Lagrangian is invariant under non-Abelian gauge transformations
\begin{equation}
A_\mu' = \Omega (A_\mu + \p_\mu) \Omega^\dagger,
\end{equation}
where $\Omega(x) \in G(2)$. Like all non-Abelian pure gauge theories, $G(2)$
Yang-Mills theory is asymptotically free. Complementary to this, at low 
energies one expects confinement.

However, in contrast to $SU(3)$ Yang-Mills theory, the triviality of the $G(2)$
center has far reaching consequences for how confinement is realized. In 
particular, as we have already seen in eq.(\ref{product}), an external static 
``quark'' in the fundamental $\{7\}$ representation can be screened by at least
three ``gluons''. Hence, via creation of dynamical ``gluons'' the confining 
string connecting two static $G(2)$ ``quarks'' can break and the potential 
flattens off. Hence, the string tension --- as the ultimate slope of the heavy 
``quark'' potential at distance $R \rightarrow \infty$ --- vanishes. Thus, in
contrast to $SU(3)$ Yang-Mills theory where gluons cannot screen quarks, there 
is a more subtle form of confinement in $G(2)$ Yang-Mills theory, very much 
like the confinement in $SU(3)$ QCD. In the QCD case screening arises due to 
dynamical quark-anti-quark pair creation, and again the confining string 
breaks. Thus, $G(2)$ pure Yang-Mills theory provides us with a suitable
theoretical laboratory to investigate confinement without facing additional
complications due to dynamical fermions. In the next section the issue of 
$G(2)$ string breaking will be studied in the strong coupling limit of lattice 
gauge theory.

Of course, unless one proves confinement analytically, one cannot be sure that
QCD, $G(2)$ Yang-Mills theory, or any other gauge theory is indeed in the
confined phase in the continuum limit. Based on general wisdom, one would
certainly expect that $G(2)$ gauge theory confines color in the same way as
QCD. In particular, we do not expect it to be in a massless non-Abelian 
Coulomb phase. An order parameter that distinguishes between a confined phase
(without a string tension, however, with color screening) and a Coulomb phase
has been constructed by Fredenhagen and Marcu \cite{Fre85}. In the next section
we will show that $G(2)$ lattice Yang-Mills theory is indeed in the confined 
phase in the strong coupling limit.

Due to the triviality of the center one also expects unusual behavior of $G(2)$
Yang-Mills theory at finite temperature. In $SU(N_c)$ Yang-Mills theory there 
is a deconfinement phase transition at finite temperature where the $\Z(N_c)$
center symmetry gets spontaneously broken. For two colors ($N_c = 2$) the 
deconfinement phase transition is second order 
\cite{McL81,Kut81,Eng81,Gav83a,Gav83b} and belongs to the universality class 
of the 3-d Ising model \cite{Eng90,Eng92}. For $N_c = 3$, on the other hand, 
the phase transition is first order 
\cite{Cel83,Kog83,Got85,Bro88,Gav89,Fuk89a,Alv90} and the bulk physics is not 
universal. This is consistent with what one expects based on the behavior of 
the 3-d 3-state Potts model \cite{Kna79,Blo79,Her79,Wu82,Fuk89b}. The 
high-temperature deconfined phase of $SU(N_c)$ Yang-Mills theory is 
characterized by a non-zero value of the Polyakov loop order parameter and by a
vanishing  string tension. On the other hand, in the low-temperature confined 
phase, the Polyakov loop vanishes and the string tension is non-zero. As we 
have seen before, already in the confined phase of $G(2)$ Yang-Mills theory the
string tension is zero. Since for $G(2)$ the center is trivial, the Polyakov 
loop no longer vanishes in the confined phase and it is hence no longer an 
order parameter for deconfinement. As a result, for $G(2)$ there is no 
compelling argument for a phase transition at finite temperature. In 
particular, a second order phase transition is practically excluded due to the 
absence of a symmetry that could break spontaneously. Even without an 
underlying symmetry, a second order phase transition can occur as an endpoint 
of a line of first order transitions. However, these particular cases require 
fine-tuning of some parameter and can thus be practically excluded in $G(2)$ 
Yang-Mills theory. On the other hand, one cannot rule out a first order phase 
transition because this does not require spontaneous symmetry breaking. Since 
the deconfinement phase transition in $SU(3)$ Yang-Mills theory is already 
rather weakly first order, we expect the $G(2)$ Yang-Mills theory to have only 
a crossover from a low- to a high-temperature regime. In QCD dynamical quarks 
also break the $\Z(3)$ center symmetry explicitly. As the quark masses are 
decreased starting from infinity, the first order phase transition of the pure 
gauge theory persists until it terminates at a critical point and then turns 
into a crossover \cite{Has83,Kar00}. Of course, in the chiral limit there is an
exact chiral symmetry that is spontaneously broken at low and restored at high 
temperatures. Hence, one expects a finite temperature chiral phase transition 
which should be second order for two and first order for three massless flavors
\cite{Pis84}. A second order chiral phase transition will be washed out to a 
crossover once non-zero quark masses are included. In $G(2)$ Yang-Mills theory 
there is no chiral symmetry that could provide us with an order parameter for a
finite temperature phase transition.

\subsection{$G(2)$ Gauge-Higgs Model}

In the next step we add a Higgs field in the fundamental $\{7\}$ representation
in order to break $G(2)$ spontaneously down to $SU(3)$. Then 6 of the 14 $G(2)$
``gluons'' pick up a mass proportional to the vacuum value $v$ of the Higgs
field, while the remaining 8 $SU(3)$ gluons are unaffected by the Higgs 
mechanism and are confined inside glueballs. For large $v$ the theory thus
reduces to $SU(3)$ Yang-Mills theory. For small $v$ (on the order of 
$\Lambda_{QCD}$), on the other hand, the additional $G(2)$ ``gluons'' are light
and participate in the dynamics. Finally, for $v = 0$ the Higgs mechanism 
disappears and we arrive at $G(2)$ Yang-Mills theory. Hence, by varying $v$ one
can interpolate smoothly between $G(2)$ and $SU(3)$ Yang-Mills theory and 
connect the exceptional $G(2)$ confinement with the usual confinement in 
$SU(3)$.

The Lagrangian of the $G(2)$ gauge-Higgs model is given by
\begin{equation}
{\cal L}_{GH}[A,\Phi] = {\cal L}_{YM}[A] + \frac{1}{2} D_\mu \Phi D_\mu \Phi + 
V(\Phi).
\end{equation}
Here $\Phi(x) = (\Phi^1(x),\Phi^2(x),...,\Phi^7(x))$ is the real-valued Higgs 
field,
\begin{equation}
D_\mu \Phi = (\p_\mu + A_\mu) \Phi,
\end{equation}
is the covariant derivative and
\begin{equation}
V(\Phi) = \lambda (\Phi^2 - v^2)^2
\end{equation}
is the scalar potential. We have seen in eq.(\ref{prod}) that the tensor 
product $\{7\} \otimes \{7\} \otimes \{7\}$ contains a singlet. Hence, one
might also expect a cubic term $T_{abc} \Phi^a \Phi^b \Phi^c$ in the 
Lagrangian. However, due to the anti-symmetry of the tensor $T$ such a term
vanishes. The product $\{7\} \otimes \{7\} \otimes \{7\} \otimes \{7\}$
contains four singlets. One corresponds to $v^2 \Phi^2$ and one to $\Phi^4$.
The other two again vanish due to antisymmetry. Hence, the scalar potential
from above is the most general one consistent with $G(2)$ symmetry and 
perturbative renormalizability.

Let us first consider the ungauged Higgs model with the Lagrangian
\begin{equation}
{\cal L}_{H}[\Phi] = \frac{1}{2} \p_\mu \Phi \p_\mu \Phi + V(\Phi).
\end{equation}
This theory has even an enlarged global $SO(7)$ symmetry which is spontaneously
broken to $SO(6)$. Due to Goldstone's theorem there are $21 - 15 = 6$ massless
bosons and one Higgs particle of mass squared $M_H^2 = 8 \lambda v^2$. When we 
now gauge only the $G(2)$ subgroup of $SO(7)$ we break the global $SO(7)$ 
symmetry explicitly. As a result, the previously intact global $SO(6) \simeq 
SU(4)$ symmetry turns into a local $SU(3)$ symmetry. Hence, a Higgs in the 
$\{7\}$ representation of $G(2)$ breaks the gauge symmetry down to $SU(3)$. The
6 massless Goldstone bosons are eaten and become the longitudinal components of
$G(2)$ ``gluons'' which pick up a mass $M_G = g v$. The remaining 8 gluons are
those familiar from $SU(3)$ Yang-Mills theory. Choosing the vacuum value of the
Higgs field as $\Phi(x) = (0,0,0,0,0,0,v)$, the unbroken $SU(3)$ invariance 
can be explicitly verified using eqs.(\ref{su3gen},\ref{g2gen}).

It is interesting to compare this situation with what happens in the standard 
model. Before gauge interactions are switched on, the standard model Higgs 
field can be viewed as a vector in the $\{4\}$ representation of $SO(4) \simeq 
SU(2)_L \otimes SU(2)_R$. When it picks up a vacuum expectation value this 
global symmetry is spontaneously broken to $SO(3) \simeq SU(2)_{L=R}$ and there
are $6 - 3 = 3$ massless Goldstone bosons. When one gauges only the $SU(2)_L 
\otimes U(1)_Y$ subgroup of $SU(2)_L \otimes SU(2)_R$ one again breaks the 
global $SO(4)$ symmetry explicitly. As a result, the previously intact global 
$SO(3) \simeq SU(2)_{L=R}$ symmetry turns into the local $U(1)_{L=R} = 
U(1)_{em}$ symmetry of electromagnetism. In this case, the 3 Goldstone bosons, 
of course, become the longitudinal components of the massive bosons $W^\pm$ and
$Z^0$.

With the Higgs mechanism in place, we can think of the $G(2)$ model from above
as an $SU(3)$ gauge theory with 6 additional vector bosons of mass $M_G$ in the
$\{3\}$ and $\{\overline 3\}$ representation and a scalar Higgs boson with mass
$M_H$ as a $\{1\}$ of $SU(3)$. Just like dynamical quarks in QCD, the massive
``gluons'' in the $\{3\}$ and $\{\overline 3\}$ representation explicitly break
the center of $SU(3)$. As a result, the confining string connecting a static 
quark-anti-quark pair can break by the creation of massive $G(2)$ ``gluons''.
As the mass of these ``gluons'' increases with $v$, the distance at which the 
string breaks becomes larger. In the limit $v \rightarrow \infty$ the 
additional ``gluons'' are removed from the theory, the $\Z(3)$ center symmetry 
is restored, and the unbreakable string of $SU(3)$ Yang-Mills theory emerges.

Using the Higgs mechanism to interpolate between $SU(3)$ and $G(2)$ Yang-Mills
theory, we again consider the issue of the deconfinement phase transition. In 
the $SU(3)$ theory this transition is weakly first order. As the mass of the 6 
additional $G(2)$ gluons is decreased, the $\Z(3)$ center symmetry of $SU(3)$
is explicitly broken and the phase transition is weakened. Qualitatively, we 
expect the heavy ``gluons'' to play a similar role as heavy quarks in $SU(3)$
QCD. Hence, we expect the first order deconfinement phase transition line to
end at a critical endpoint before the additional $G(2)$ ``gluons'' have become
massless \cite{Has83,Kar00}. In that case, the pure $G(2)$ Yang-Mills theory 
should have no deconfinement phase transition, but merely a crossover. 

\subsection{$G(2)$ QCD}

Let us now consider $G(2)$ gauge theory with $N_f$ flavors of fermions. As
before, we will use the Higgs mechanism induced by a scalar field in the 
$\{7\}$ representation to interpolate between $G(2)$ and $SU(3)$ QCD. We 
introduce $G(2)$ ``quarks'' as Majorana fermions in the fundamental 
representation. Since all $G(2)$ representations are real, a Dirac fermion 
simply represents a pair of Majorana fermions. Hence, it is most natural to 
work with Majorana ``quarks'' as fundamental objects. Under $SU(3)$ subgroup 
transformations a $\{7\}$ of $G(2)$ decomposes into $\{3\} \oplus 
\{\overline 3\} \oplus \{1\} $. Hence, when $G(2)$ is broken down to $SU(3)$, a
$\{7\}$ Majorana ``quark'' of $G(2)$ turns into an ordinary Dirac quark $\{3\}$
and its anti-quark $\{\overline 3\}$ as well as a color singlet Majorana 
fermion that does not participate in the strong interactions. The $G(2)$ 
Majorana ``quark'' spinor can be written as
\begin{equation}
\lambda = \left(\begin{array}{c} \Psi \\ C \overline \Psi^T \\ \chi \end{array}
\right),
\end{equation}
where $\Psi$ is an $SU(3)$ Dirac quark spinor, $\chi$ is the color singlet 
Majorana fermion, and $C$ is the charge-conjugation matrix. The ``anti-quark'' 
spinor $\overline \lambda$ is related to $\lambda$ by charge-conjugation
\begin{equation}
\overline \lambda = (\overline \Psi,- \Psi^T C^{-1},- \chi^T C^{-1}).
\end{equation}
The Lagrangian of $G(2)$ QCD takes the form
\begin{equation}
{\cal L}_{QCD}[A,\overline \lambda,\lambda] = {\cal L}_{YM}[A] + 
\frac{1}{2} \overline \lambda \gamma_\mu (\p_\mu + A_\mu) \lambda.
\end{equation}
In $G(2)$ gauge theory, quark masses arise from Yukawa couplings to the scalar 
field as well as from Majorana mass terms. For simplicity, in what follows we 
consider massless ``quarks'' only. 

\subsubsection{The $N_f = 1$ Case}

As a first step we consider a single flavor --- say the $u$-quark. Ordinary 
$N_f = 1$ $SU(3)$ QCD has a $U(1)_B$ symmetry --- just baryon number which is 
unbroken. In particular, there are no massless Goldstone bosons and the theory 
has a mass-gap. Color singlet states include $u \overline u$ mesons with a 
valance quark and anti-quark as well as a $uuu$ baryon $\Delta^{++}$ with three
valance quarks. The lightest particle in this theory is presumably a 
vector-meson similar to the physical $\omega$-meson. Like in ordinary QCD, the 
pseudo-scalar meson $\eta'$ gets its mass via the anomaly from topological 
charge fluctuations. Only in the large $N_c$ limit it becomes a Goldstone 
boson. For $N_c = 3$ it may or may not be lighter than the vector-meson.

As we have seen before, in $G(2)$ gauge theory ``quarks'' and ``anti-quarks'' 
are indistinguishable. Consequently, the $U(1)_{L=R} = U(1)_B$ baryon number 
symmetry of $SU(3)$ QCD is reduced to a $\Z(2)_B$ symmetry. One can only 
distinguish between states with an even and odd number of ``quark'' 
constituents. In particular, eq.(\ref{product}) implies that one can construct 
a colorless state $uGGG$ with one $G(2)$ ``quark'' screened by three $G(2)$ 
``gluons''. This state mixes with other states containing an odd number of
quarks --- e.g. with the usual $uuu$ states --- to form the $G(2)$ ``baryon''. 
In contrast to $SU(3)$ QCD, two $G(2)$ ``baryons'' (which are odd under
$\Z(2)_B$) can annihilate into ``mesons''. When one uses the Higgs mechanism to
break $G(2)$ to $SU(3)$, one can remove the 6 additional $G(2)$ ``gluons'' by 
increasing the vacuum value $v$. As a consequence, the states $uGGG$ become 
heavy and can no longer mix with $uuu$. As a result, the standard $U(1)_B$ 
baryon number symmetry of $SU(3)$ QCD emerges as an approximate symmetry. As 
long as $v$ remains finite, the heavy $G(2)$ ``gluons'' mediate weak baryon 
number violating processes. Only in the limit $v \rightarrow \infty$ $U(1)_B$ 
becomes an exact symmetry.

\subsubsection{The $N_f \geq 2$ Case}

Let us first remind ourselves of standard two flavor QCD with $SU(3)$ color 
gauge group. The chiral symmetry then is $SU(2)_L \otimes SU(2)_R \otimes
U(1)_B$ which is spontaneously broken to $SU(2)_{L=R} \otimes U(1)_B$. 
Consequently, there are $7 - 4 = 3$ massless Goldstone pions --- $\pi^+$, 
$\pi^0$ and $\pi^-$. For a $G(2)$ Majorana ``quark'' left- and right-handed 
components cannot be rotated independently by unitary transformations 
$L \in SU(2)_L$ and $R \in SU(2)_R$. In fact, the Majorana condition requires 
$L = R^*$. Hence, the chiral symmetry of $N_f = 2$ $G(2)$ QCD is 
$SU(2)_{L=R^*} \otimes \Z(2)_B$. Note that in the same way $U(1)_B = 
U(1)_{L=R}$ is reduced to $U(1)_{L=R^*=R} = \Z(2)_B$. The reduced chiral 
symmetry of $G(2)$ QCD is expected to still break spontaneously to the maximal 
vector subgroup which is now $SU(2)_{L=R^*=R} \otimes \Z(2)_B = SO(2)_{L=R} 
\otimes \Z(2)_B$. In this case, there are only $3 - 1 = 2$ Goldstone bosons. 
They can be identified as $\pi^0$ and the linear combination of $\pi^+$ and 
$\pi^-$ that is even under charge-conjugation. The mixing between $\pi^+$ and 
$\pi^-$ is induced by the exchange of one of the 6 $G(2)$ ``gluons'' that do 
not belong to $SU(3)$. The linear combination of $\pi^+$ and $\pi^-$ that is 
odd under charge-conjugation has a non-zero mass and is not a Goldstone boson 
of $G(2)$ QCD. When we remove the 6 additional ``gluons'' via the Higgs 
mechanism, the mixing of $\pi^+$ and $\pi^-$ becomes weaker as $v$ increases. 
Consequently, the mass splitting between the charge-conjugation even and odd 
states also decreases until it ultimately vanishes at $v = \infty$. In this 
limit the larger chiral symmetry of $SU(3)$ QCD emerges and we are left with
three massless pions.

It is straightforward to generalize the above discussion to arbitrary 
$N_f \geq 2$. For $SU(3)$ QCD with general $N_f$ the chiral symmetry is
$SU(N_f)_L \otimes SU(N_f)_R \otimes U(1)_B$ which is spontaneously broken to 
$SU(N_f)_{L=R} \otimes U(1)_B$, and there are $N_f^2 - 1$ massless Goldstone 
bosons. As before, the Majorana condition requires $L = R^*$. Hence, the chiral
symmetry of $G(2)$ QCD with $N_f$ massless Majorana ``quarks'' is 
$SU(N_f)_{L=R^*} \otimes \Z(2)_B$, which is expected to break spontaneously to 
$SU(N_f)_{L=R^*=R} \otimes \Z(2)_B = SO(N_f)_{L=R} \otimes \Z(2)_B$. Then there
are only $N_f(N_f + 1)/2 - 1$ Goldstone bosons. These consist of $N_f - 1$ 
neutral Goldstone bosons ($\pi^0$ and $\eta$ for $N_f =3$) and $N_f(N_f - 1)/2$
charge-conjugation even combinations of charged states ($\pi^+ + \pi^-$, 
$K^+ + K^-$, $K^0 + \overline{K^0}$ for $N_f =3$).

\subsection{Supersymmetric $G(2)$ Yang-Mills Theory}

In this section we add a single flavor adjoint Majorana gluino $\lambda$ to
the pure gluon Yang-Mills theory and thus turn it into ${\cal N} = 1$ 
supersymmetric Yang-Mills theory. First, we compare the $SU(3)$ to the $G(2)$ 
case.

Let us first describe the situation in $SU(3)$ supersymmetric Yang-Mills 
theory. Then, in addition to the gluons, there is a color octet of Majorana
gluinos. The chiral symmetry of this theory is $\Z(3)_\chi \otimes \Z(2)_B$
where $\Z(2)_B$ is the fermion number symmetry of the Majorana fermion, and 
$\Z(3)_\chi$ is a remnant of the axial $U(1)_R$ symmetry which is broken by the
anomaly. At low temperatures the $\Z(3)_\chi$ symmetry is spontaneously broken
through the dynamical generation of a gluino condensate 
$\langle \lambda \lambda \rangle$. Since both gluons and gluinos are 
in the adjoint representation, the $\Z(3)$ center of the color gauge group is
an exact symmetry of supersymmetric $SU(3)$ Yang-Mills theory. At low
temperature this discrete symmetry is unbroken, just as in the 
non-supersymmetric case. As a result, external static quarks and anti-quarks
are confined to one another through an unbreakable color flux string. 

As a consequence of the spontaneous breakdown of the discrete $\Z(3)_\chi$ 
symmetry, there are, in fact, three different low-temperature confined phases
which are distinguished by the $\Z(3)_\chi$ phase of the gluino condensate.
When such phases coexist with one another, they are separated by 
confined-confined domain walls with a non-zero interface tension. These walls
are topological defects which are characterized by the zeroth homotopy group
$\Pi_0[\Z(3)_\chi] = \Z(3)$. Based on M-theory, Witten has argued that the 
walls behave like D-branes and the confining string (which behaves like a 
fundamental string) can end on the walls \cite{Wit97}. In a field theoretical 
context this phenomenon has been explained in \cite{Cam98}. Just like other 
topological excitations, such as monopoles, cosmic strings and vortices, a 
supersymmetric domain wall has the unbroken symmetry phase in its core. 
Consequently, inside the wall (as well as inside the string) the chiral 
$\Z(3)_\chi$ symmetry is restored. Interestingly, the restoration of 
$\Z(3)_\chi$ induces the spontaneous breakdown of the center symmetry $\Z(3)$. 
Hence, the core of a supersymmetric domain wall is in the deconfined phase. 
When a confining string enters the wall the color flux that it carries spreads 
out and the string ends.

As one increases the temperature, a transition to a deconfined phase with 
restored chiral symmetry occurs. As usual, in the deconfined phase the $\Z(3)$ 
center symmetry is spontaneously broken. When a confined-confined domain wall 
is heated up to the phase transition, the deconfined phase in its core expands 
and forms a complete wetting layer whose width diverges at the phase transition
\cite{Cam98}. Due to the broken $\Z(3)$ center symmetry there are also three 
distinct deconfined phases. When those coexist, they are separated by 
deconfined-deconfined domain walls. As the phase transition is approached from 
above, similarly, a deconfined-deconfined domain wall splits into a pair of 
confined-deconfined interfaces and its core turns into a complete wetting layer
of confined phase \cite{Fre89,Tra92}.

Let us now ask how the situation is modified for $G(2)$ supersymmetric 
Yang-Mills theory for which the center is trivial. Interestingly, the remnant
chiral symmetry is enhanced to $\Z(4)_\chi \otimes \Z(2)_B$ which breaks
spontaneously to $\Z(2)_B$. As a result, similar to the $SU(3)$ case, there are
now four different low-temperature chirally broken phases which are 
characterized by the $\Z(4)_\chi$ phase $\pm 1, \pm i$ of the ``gluino'' 
condensate. Due to the triviality of the center, we have the exceptional 
confinement with a breakable string that we have already discussed in the 
non-supersymmetric $G(2)$ Yang-Mills theory. When two distinct chirally broken 
phases coexist, they are again separated by a domain wall. In contrast to the 
$SU(3)$ case, $G(2)$ strings can not only end on such walls because they can 
break and thus end anywhere.

When $G(2)$ ``gluons'' and ``gluinos'' are heated up, their chiral symmetry is
restored in a finite temperature phase transition. In contrast to $G(2)$ pure
Yang-Mills theory, a phase transition must exist because there is now an exact
spontaneously broken $\Z(4)_\chi$ chiral symmetry for which the ``gluino'' 
condensate provides us with an order parameter. As another consequence of the 
triviality of the center, there is only one high-temperature chirally symmetric
phase. In particular, deconfined-deconfined domain walls do not exist. However,
there are now two types of domain walls in the low-temperature phase. A wall of
type I separates chirally broken phases whose ``gluino'' condensates are 
related by a $\Z(4)_\chi$ transformation $- 1$, while for a wall of type II 
the phases are related by a $\pm i$-rotation. The chiral phase transition may 
be first or second order. In the latter case the low- and high-temperature 
phases do not coexist at the phase transition and complete wetting does not 
arise. However, we find it more natural to expect a first order phase 
transition. For example, the deconfinement phase transition of an ordinary 
$SU(4)$ Yang-Mills theory, which has a $\Z(4)$ center symmetry, is first order 
\cite{Whe84,Goc84,Bat84,Win01,Tep02}. If the chiral phase transition of $G(2)$ 
supersymmetric Yang-Mills theory is first order as well, the low- and 
high-temperature phases coexist and complete wetting may arise. When a wall of 
type I is heated up to the phase transition, we expect it to split into a pair 
of interfaces with a complete wetting layer of chirally symmetric phase in 
between. It is less clear if complete wetting would also occur for domain walls
of type II. We do not enter this discussion here. In any case, complete wetting
is no longer needed for strings to end on the walls.

\section{$G(2)$ Lattice Gauge Theory at Strong Coupling}

In order to substantiate some of the claims made in the previous sections we 
now formulate $G(2)$ Yang-Mills theory on the lattice and derive some analytic
results in the strong coupling limit. As usual, such results do not directly 
apply to the continuum limit and should ultimately be extended by Monte Carlo
simulations into the weak coupling continuum regime. Still, assuming that there
is no phase transition separating the strong from the weak coupling regime, the
strong coupling results provide insight into dynamical behavior --- such as 
confinement --- which persists in the continuum limit. For example, for $G(2)$
--- in agreement with the expectations --- the lattice strong coupling 
expansion confirms that the color flux string can break by dynamical ``gluon'' 
creation. In this sense, the string tension is zero and the Wilson loop is no 
longer a good order parameter. In order to characterize the phase of the 
theory, in particular, in order to distinguish between a massive confinement 
phase like in full QCD and a massless Coulomb phase, one can use the 
Fredenhagen-Marcu order parameter \cite{Fre85} which we calculate analytically 
at strong coupling. Indeed, this confirms that $G(2)$ lattice Yang-Mills theory
confines color in the same way as $SU(3)$ QCD.

The construction of $G(2)$ Yang-Mills theory on the lattice follows the 
standard procedure. The link matrices $U_{x,\mu} \in G(2)$ are group elements 
in the fundamental $\{7\}$ representation, i.e. they can be chosen entirely 
real. The standard Wilson plaquette action takes the usual form
\begin{equation}
S[U] = - \frac{1}{g^2} \sum_\Box \mbox{Tr} \ U_\Box =
- \frac{1}{g^2} \sum_{x,\mu < \nu} \mbox{Tr} \
U_{x,\mu} U_{x+\hat\mu,\nu} U^\dagger_{x+\hat\nu,\mu} U^\dagger_{x,\nu},
\end{equation}
where $g$ is the bare gauge coupling. The partition function is given by
\begin{equation}
Z = \int {\cal D}U \exp(- S[U]),
\end{equation}
where the measure of the path integral
\begin{equation}
\int {\cal D}U = \prod_{x,\mu} \int_{G(2)} dU_{x,\mu},
\end{equation}
is a product of local Haar measures of the group $G(2)$ for each link. By
construction, both the action and the measure are explicitly invariant under
gauge transformations
\begin{equation}
U'_{x,\mu} = \Omega_x U_{x,\mu} \Omega^\dagger_{x+\hat\mu},
\end{equation}
with $\Omega_x \in G(2)$. The Wilson loop
\begin{equation}
W_{\cal C} = \mbox{Tr} \ {\cal P} \prod_{(x,\mu) \in {\cal C}} U_{x,\mu}
\end{equation}
is the trace of a path ordered product of link variables along the closed path
${\cal C}$, and its expectation value is given by
\begin{equation}
\langle W_{\cal C} \rangle = 
\frac{1}{Z} \int {\cal D}U \ W_{\cal C} \exp(- S[U]).
\end{equation}
For a rectangular path ${\cal C}$ of extent $R$ in the spatial and $T$ in the
temporal direction the Wilson loop 
\begin{equation}
\lim_{T \rightarrow \infty} \langle W_{\cal C} \rangle = \exp(- V(R) T)
\end{equation}
determines the potential $V(R)$ between static color sources at distance $R$.
In a phase with a linearly rising confining potential $V(R) \sim \sigma R$,
where $\sigma$ is the string tension, the Wilson loop obeys an area law. If 
the potential levels off at large distances, the Wilson loop obeys a perimeter 
law. In the strong coupling limit of $SU(3)$ Yang-Mills theory the Wilson loop 
indeed follows an area law. In $G(2)$ Yang-Mills theory, on the other hand, we 
expect static ``quarks'' to be screened by ``gluons'' and hence the string to 
break. As a result, the static ``quark'' potential levels off and large Wilson 
loops obey a perimeter law.

In the strong coupling regime $g^2 \gg 1$, and $1/g^2$ can be used as a small 
expansion parameter. The first step of the strong coupling expansion is the 
character expansion of the Boltzmann factor for an individual plaquette
\begin{equation}
\exp(\frac{1}{g^2} \mbox{Tr} \ U_\Box) = 
\sum_\Gamma c_\Gamma(\frac{1}{g^2}) \chi_\Gamma(U_\Box),
\end{equation}
where $\Gamma$ is a generic representation of the gauge group and the
corresponding character $\chi_\Gamma(U_\Box)$ is the trace of the matrix 
$U_\Box$ in that representation. The coefficients $c_\Gamma(1/g^2)$ enter the 
strong coupling expansion as power-series in $1/g^2$. For example, for $G(2)$ 
we have
\begin{eqnarray}
c_{\{1\}}(\frac{1}{g^2})&=&
1 + \frac{1}{2 g^4} + \frac{1}{6 g^6} + \frac{1}{6 g^8} + 
\frac{1}{12 g^{10}} + \frac{7}{144 g^{12}} + ... , \nonumber \\
c_{\{7\}}(\frac{1}{g^2})&=&
\frac{1}{g^2} + \frac{1}{2 g^4} + \frac{2}{3 g^6} + 
\frac{5}{12 g^8} + \frac{7}{24 g^{10}} + \frac{1}{6 g^{12}} + ... , 
\nonumber \\
c_{\{14\}}(\frac{1}{g^2})&=&
\frac{1}{2 g^4} + \frac{1}{3 g^6} + \frac{3}{8 g^8} + 
\frac{1}{4 g^{10}} + \frac{1}{6 g^{12}} + ... .
\end{eqnarray}

Let us now compute the expectation value of a rectangular Wilson loop of size
$R \times T$ in the strong coupling limit. In $SU(N_c)$ lattice Yang-Mills 
theory the leading contribution in the strong coupling expansion results from 
tiling the rectangular surface enclosed by the Wilson loop with $R \times T$ 
elementary plaquettes in the fundamental representation. This gives rise to the
strong coupling area law. All higher order contributions amount to deformations
of this minimal surface of plaquettes. Such contributions are also present for 
$G(2)$. However, in the $G(2)$ case there are additional terms arising from a 
tube of plaquettes along the perimeter of the Wilson loop. In fact, these 
contributions dominate at large $R$ and give rise to a strong coupling 
perimeter law. For small $R$, on the other hand, the surface term dominates and
yields a linearly rising potential at short distances. It should be noted that 
tube contributions arise even in $SU(N_c)$ Yang-Mills theory for Wilson loops 
of adjoint charges. In that case, there is again no linearly rising confinement
potential. Due to the triviality of the center, for $G(2)$ the perimeter law
arises already for fundamental charges.

Here we consider only the leading contribution to the strong coupling 
expansion. For small $R \leq R_c$ the surface term dominates and gives
\begin{equation}
\langle W_{\cal C} \rangle = 7 (\frac{1}{7 g^2})^{RT} 
\stackrel{T \rightarrow \infty}\longrightarrow \exp(- V(R) T),
\end{equation}
which yields a linear potential
\begin{equation}
V(R) = - \log(\frac{1}{7 g^2}) R.
\end{equation}
In an $SU(N_c)$ Yang-Mills theory the linear potential would extend to 
arbitrary distances and $- \log(1/7 g^2)$ would play the role of the string 
tension. In the $G(2)$ case, however, the large $R \geq R_c$ behavior is 
dominated by the perimeter term
\begin{equation}
\langle W_{\cal C} \rangle = 4 (\frac{1}{7g^2})^{8 (R + T -2)},
\end{equation}
which gives rise to a flat ($R$-independent) potential
\begin{equation}
V(R) = - 8 \log(\frac{1}{7 g^2}).
\end{equation}
At distances larger than $R_c = 8$ the perimeter term overwhelms the surface 
term and the string breaks. Hence, there is no confinement in the sense of a
non-vanishing string tension characterizing the slope of the potential at 
asymptotic distances. When one pulls apart a $G(2)$ ``quark'' pair beyond the 
distance $R_c$, ``gluons'' pop up from the vacuum and screen the fundamental 
color charges of the ``quarks''. This is possible only because $G(2)$ has a 
trivial center. From the tensor product decomposition of eq.(\ref{product}) one
infers that at least three ``gluons'' (which are in the $\{14\}$ of $G(2)$) are
needed to screen a single ``quark'' (in the $\{7\}$ of $G(2)$).

Since its string can break, pure $G(2)$ Yang-Mills theory resembles full QCD.
In that case, dynamical quark-anti-quark pairs materialize from the vacuum to
screen an external static quark-anti-quark pair at large separation. Hence, 
also in full QCD the static quark-anti-quark potential flattens off and 
ultimately there is no string tension. Of course, this does not mean that QCD
does not confine. In particular, there should be no single quark or gluon 
states in the physical spectrum, i.e. QCD should not be realized in a 
non-Abelian Coulomb phase. An order parameter that distinguishes between 
Coulomb and confinement phases (even if there is no string tension) has been
constructed by Fredenhagen and Marcu \cite{Fre85}. This order parameter can
also be adapted to $G(2)$ pure Yang-Mills theory and can, in fact, be evaluated
in the strong coupling limit.

The Fredenhagen-Marcu order parameter is a ratio of two expectation values. The
numerator consists of parallel transporters along an open staple-shaped path 
connecting a source and a sink of color flux and the denominator is the square 
root of a closed Wilson loop
\begin{equation}
\rho(R,T) = \frac{\langle 
\input{FredMarcu_num.pictex}
\rangle}{\;\;\;\langle 
\input{FredMarcu_den.pictex}
\rangle^{1/2} }.
\end{equation}
The open path symbolic object in the numerator stands for
\begin{equation}
\input{FredMarcu_num.pictex}
 = \mbox{Tr}(U_{\Box_x} \Lambda_a) 
\mbox{Tr}[\Lambda_a U_{{\cal C}_{xy}} \Lambda_b U_{{\cal C}_{xy}}^\dagger]
\mbox{Tr}(U_{\Box_y}^\dagger \Lambda_b),
\end{equation}
where
\begin{equation}
U_{{\cal C}_{xy}} = {\cal P} \prod_{(z,\mu) \in {\cal C}_{xy}} U_{z,\mu}
\end{equation}
is a path-ordered product of parallel transporters along the open path 
${\cal C}_{xy}$. This staple-shaped path of time-extent $T$ connects the source
and sink points $x$ and $y$ that are spatially separated by a distance $R$. At 
these points dynamical $G(2)$ ``gluons'' are created by plaquette operators 
$U_{\Box_x}$ and $U_{\Box_y}$. The factors $\Lambda_a$ and $\Lambda_b$ reflect 
the fact that ``gluons'' transform in the adjoint representation. The closed 
path symbolic object in the denominator is a Wilson loop of size $R \times 2 T$
in the adjoint representation.

The Fredenhagen-Marcu order parameter describes the creation of a pair of 
adjoint dynamical charges that propagate for a time $T$ and measures their 
overlap with the vacuum in the limit $R, T \rightarrow \infty$. In a Coulomb 
phase charged states exist in the physical spectrum and are orthogonal to the 
vacuum. Consequently, the Fredenhagen-Marcu vacuum overlap order parameter then
vanishes. In a confined phase, on the other hand, the dynamical charges are 
screened and $\rho(R,T)$ goes to a non-zero constant for large $R$ and $T$.

In the strong coupling limit the leading contribution to the numerator of the
vacuum overlap order parameter is a tube of plaquettes emanating from the 
source plaquette $\Box_x$, following the staple-shaped path, and ending at the 
sink plaquette $\Box_y$. This leads to a perimeter law in the numerator. Just
like the Wilson loop in the fundamental representation that was calculated 
before, the adjoint Wilson loop in the denominator of the order parameter also
obeys a perimeter law. Due to the square root and the doubled temporal extent,
the perimeter behavior in the numerator and the denominator cancel exactly and
one is left with
\begin{equation}
\label{Fredord}
\rho(R,T) = \frac{112 (1/7 g^2)^{4(2T+R)}}
{2 (1/7 g^2)^{4(2T+R-2)}} = 56 (\frac{1}{7 g^2})^8
\end{equation}
Here the plane of the plaquettes $\Box_x$ and $\Box_y$ is dual to the plane of 
the staple-shaped path ${\cal C}_{xy}$. Eq.(\ref{Fredord}) shows that we are 
indeed in a confined phase (without a string tension, however, with color 
charge screening) and not in a non-Abelian Coulomb phase. Of course, this 
strong coupling result does not guarantee that $G(2)$ Yang-Mills theory 
confines also in the continuum limit, as one would naturally expect. It would 
be interesting to investigate this issue in numerical simulations.

One might argue that in a pure gluon theory quarks are simply not present and 
can hence not even be used as external static sources. If one wants to study 
confinement of gluons without using static quarks, one can use the vacuum
overlap operator also in an $SU(3)$ Yang-Mills theory. In the strong coupling 
limit one then finds again that gluons are in a confined phase with color 
screening by dynamical gluon creation --- and are not in a Coulomb phase.

It should be noted that the Fredenhagen-Marcu order parameter makes sense only
at zero temperature, because it requires to take the limit $T \rightarrow 
\infty$. Since $G(2)$ has a trivial center (and thus a vanishing string 
tension) there is no need for the standard finite temperature deconfinement
phase transition. In particular, there is no center symmetry that could break
spontaneously at high temperatures. Of course, this argument does not exclude
the existence of a first order phase transition at finite temperature. We find
it more natural to expect just a crossover. Again, this is an interesting point
for numerical investigation. In any case, analytic strong coupling calculations
cannot answer this question.

\section{Conclusions}

We have compared qualitative non-perturbative features such as confinement and 
chiral symmetry breaking in theories with $G(2)$ and $SU(3)$ gauge groups. In
particular, we have exploited the Higgs mechanism (induced by a scalar field in
the $\{7\}$ representation of $G(2)$) in order to interpolate smoothly between
these two cases. We have focused on effects which are intimately related to the
center of the gauge group, and which hence are qualitatively different for
$SU(3)$ with center $\Z(3)$ and $G(2)$ with a trivial center.

When all dynamical fields in an $SU(3)$ gauge theory are center-blind (such as
gluons or gluinos which have trivial triality) the $\Z(3)$ center is an exact
symmetry. Infinitely heavy quarks with non-trivial triality can be used as 
external probes of the gluon dynamics that provide information about how the 
$\Z(3)$ symmetry is realized. In a confined phase with intact center symmetry
the confining string connecting a static quark-anti-quark pair is absolutely
unbreakable and has a non-zero string tension that characterizes the 
interaction at arbitrarily large distances. The string tension can vanish only
when the center gets spontaneously broken, which is indeed unavoidable at high
temperatures. Then the Euclidean time extent is short and the gauge field 
configuration becomes almost static. As a consequence, the Polyakov loop order 
parameter becomes non-zero. Hence, the exact center symmetry provides us with 
an argument for the existence of a deconfinement phase transition. If the
transition is second order, universality arguments suggest that it is in the
universality class of a 3-d center-symmetric scalar field theory for the 
Polyakov loop \cite{Sve82}. For example, for $N_c = 2$ it is second 
order \cite{McL81,Kut81,Eng81,Gav83a,Gav83b} and falls in the universality 
class of the 3-d Ising model \cite{Eng90,Eng92}.

Since it has a trivial center, the concept of triality does not extend to 
$G(2)$. Consequently, any infinitely heavy external source can be screened by
dynamical ``gluons'' and thus the string always breaks at large distances 
through the creation of dynamical ``gluons''. As a result, the string tension 
ultimately vanishes. However, a strong coupling lattice study of the 
Fredenhagen-Marcu vacuum overlap order parameter shows that the 
theory is still confining --- i.e. no colored states exist in the spectrum.
Confinement without a (fundamental) string tension is indeed exceptional for a 
pure gauge theory. It only arises for the exceptional Lie groups $G(2)$, $F(4)$
and $E(8)$. As another consequence of the trivial center, the $G(2)$ Polyakov 
loop is no longer an order parameter. Hence, in contrast to $SU(N_c)$ 
Yang-Mills theory, for $G(2)$ there is no compelling reason for a finite 
temperature deconfinement phase transition. We cannot exclude a first order 
phase transition but we expect only a crossover. Clearly, the triviality of the
center implies less predictive power about a possible phase transition. 

Once dynamical fields with non-trivial triality (such as light quarks) are 
included in an $SU(3)$ gauge theory, they break the center symmetry explicitly.
As a result, the string connecting a static quark-anti-quark pair can now break
through pair creation of dynamical quarks and the string tension ultimately 
vanishes. Again, the Fredenhagen-Marcu order parameter still signals
confinement. In addition, the Polyakov loop is no longer a good order 
parameter. From this point of view, the confinement in $G(2)$ Yang-Mills theory
resembles the one of $SU(3)$ QCD, and hence, it is not so exceptional after 
all.

Again, by using the Higgs mechanism, we have also interpolated between $G(2)$ 
and $SU(3)$ gauge theories with massless dynamical fermions, both in the 
fundamental and in the adjoint representation. In many of these cases, there is
a non-trivial chiral symmetry that breaks spontaneously at low temperatures.
Since the $G(2)$ representations are real, we have considered $N_f$ flavors of
Majorana fermions. The chiral symmetry then is $SU(N_f)_{L=R^*} \otimes 
\Z(2)_B$ which breaks spontaneously to $SO(N_f)_{L=R} \otimes \Z(2)_B$. It is
interesting how this pattern of symmetry breaking turns into the breaking of 
$SU(N_f)_L \otimes SU(N_f)_R \otimes U(1)_B$ to $SU(N_f)_{L=R} \otimes U(1)_B$
that occurs in QCD.

As we have seen, there are many interesting non-perturbative phenomena that
arise in $G(2)$ gauge theories. Despite the fact that Nature chose not to use
$G(2)$ (at least at presently accessible energies) it may be of theoretical 
interest to study $G(2)$ gauge theories more quantitatively. Lattice gauge
theory provides us with a powerful tool for such investigations. For example,
it would be interesting to check if the strong coupling confined phase extends
to the continuum limit, by measuring the Fredenhagen-Marcu order parameter in a
numerical simulation. With lattice methods one can also decide if the low- and
high-temperature regimes in $G(2)$ Yang-Mills theory are separated by a first
order phase transition or just by a crossover. 

It is also interesting to investigate Yang-Mills theories with other gauge 
groups such as $Sp(N)$, which have a $\Z(2)$ center symmetry. If they have a
second order deconfinement phase transition, one expects it to be in the
universality class of the 3-d Ising model. A numerical study of $Sp(2)$ gauge 
theory is presently in progress \cite{Hol03}. The group $Sp(2)$ with 10 
generators is the fourth of the rank 2 Lie groups besides $SO(4) \simeq SU(2)
\otimes SU(2)$, $SU(3)$ and $G(2)$. Based on its rank and its number of 
generators one might expect that it should behave more like $SU(3)$ than like 
$SU(2) = Sp(1)$. However, as in the $SU(2)$ case, we find that $Sp(2)$
Yang-Mills theory has a second order deconfinement phase transition with 3-d 
Ising critical exponents.

In conclusion, we have used $G(2)$ gauge theories as a theoretical laboratory
to study $SU(3)$ theories in an unusual environment. In particular, the
embedding of $SU(3)$ in $G(2)$ with its trivial center forces us to think about
confinement without the luxury of the $\Z(3)$ symmetry. As one would expect,
confinement itself works perfectly well without the center symmetry. However,
in its absence we loose predictive power about a possible phase transition at
finite temperature. 

\section*{Acknowledgments}

We like to thank S.~Caracciolo, P.~de~Forcrand, R.~Ferrari, O.~Jahn, J.~Kuti 
and  A.~Smilga for interesting discussions. This work is supported by the DOE
under the grant DOE-FG03-97ER40546, by the Schweizerischer Nationalfond, as 
well as by the European Community's Human Potential Program under the grant
HPRN-CT-2000-00145 Hadrons/Lattice QCD, BBW Nr. 99.0143.

\begin{appendix}

\section{Transition Functions, Twist-Tensor and Consistency Conditions}

\label{ch:appC}

In this appendix we derive some relations for periodic and $C$-periodic 
boundary conditions that are used in section 2. First, we consider periodic
boundary conditions. Shifting the gauge field in two orthogonal directions 
$e_\nu$ and $e_\rho$, on the one hand, one obtains
\begin{eqnarray}
\!\!\!\!\!A_\mu(x + L_\nu e_\nu + L_\rho e_\rho)&=& 
\Omega_\nu(x + L_\rho e_\rho) (A_\mu(x + L_\rho e_\rho) + \p_\mu) 
\Omega_\nu(x + L_\rho e_\rho)^\dagger \nonumber \\
&=&\Omega_\nu(x + L_\rho e_\rho) \Omega_\rho(x) (A_\mu(x) + \p_\mu) 
\Omega_\rho(x)^\dagger \Omega_\nu(x + L_\rho e_\rho)^\dagger. \nonumber \\ \,
\end{eqnarray}
On the other hand, by performing the two shifts in the opposite order, one 
finds
\begin{eqnarray}
\!\!\!\!\!A_\mu(x + L_\rho e_\rho + L_\nu e_\nu)&=& 
\Omega_\rho(x + L_\nu e_\nu) (A_\mu(x + L_\nu e_\nu) + \p_\mu) 
\Omega_\rho(x + L_\nu e_\nu)^\dagger \nonumber \\
&=&\Omega_\rho(x + L_\nu e_\nu) \Omega_\nu(x) (A_\mu(x) + \p_\mu) 
\Omega_\nu(x)^\dagger \Omega_\rho(x + L_\nu e_\nu)^\dagger. \nonumber \\ \,
\end{eqnarray}
The two results are consistent only if the transition functions obey the 
cocycle condition eq.(\ref{cocycle}).

Eq.(\ref{transition}) guarantees gauge-covariance of the boundary condition, 
i.e.
\begin{eqnarray}
A_\mu(x + L_\nu e_\nu)'&=&\Omega(x + L_\nu e_\nu) 
(A_\mu(x + L_\nu e_\nu) + \p_\mu) \Omega(x + L_\nu e_\nu)^\dagger 
\nonumber \\
&=&\Omega(x + L_\nu e_\nu) \Omega_\nu(x) (A_\mu(x) + \p_\mu) 
\Omega_\nu(x)^\dagger \Omega(x + L_\nu e_\nu)^\dagger \nonumber \\
&=&\Omega_\nu(x)' (A_\mu(x)' + \p_\mu) {\Omega_\nu(x)'}^\dagger.
\end{eqnarray}
Interestingly, the gauge transformed cocycle condition takes the form
\begin{eqnarray}
&&\Omega_\nu(x + L_\rho e_\rho)' \Omega_\rho(x)' = 
\Omega_\rho(x + L_\nu e_\nu)' \Omega_\nu(x)' z_{\nu\rho}' \ \Rightarrow
\nonumber \\
&&\Omega(x + L_\rho e_\rho + L_\nu e_\nu)
\Omega_\nu(x + L_\rho e_\rho) \Omega(x + L_\rho e_\rho)^\dagger 
\Omega(x + L_\rho e_\rho) \Omega_\rho(x) \Omega(x)^\dagger = 
\nonumber \\
&&\Omega(x + L_\nu e_\nu + L_\rho e_\rho) 
\Omega_\rho(x + L_\nu e_\nu) \Omega(x + L_\nu e_\nu)^\dagger 
\Omega(x + L_\nu e_\nu) \Omega_\nu(x) \Omega(x)^\dagger z_{\nu\rho}' \
\Rightarrow \nonumber \\
&&\Omega_\nu(x + L_\rho e_\rho) \Omega_\rho(x) = 
\Omega_\rho(x + L_\nu e_\nu) \Omega_\nu(x) z_{\nu\rho}'.
\end{eqnarray}
Consequently, $z_{\nu\rho}' = z_{\nu\rho}$, i.e. the twist-tensor is gauge
invariant.

Next, we consider $C$-periodic boundary conditions. As before, we shift the 
gauge field in two orthogonal directions. First, we pick two different spatial 
directions $i$ and $j$, and we obtain
\begin{eqnarray}
\!\!\!\!\!A_\mu(x + L_i e_i + L_j e_j)&=& 
\Omega_i(x + L_j e_j) (A_\mu(x + L_j e_j)^* + \p_\mu) 
\Omega_i(x + L_j e_j)^\dagger \nonumber \\
&=&\Omega_i(x + L_j e_j) \Omega_j(x)^* (A_\mu(x) + \p_\mu) 
\Omega_j(x)^T \Omega_i(x + L_j e_j)^\dagger. \nonumber \\ \,
\end{eqnarray}
Performing the two shifts in the opposite order, one now finds
\begin{eqnarray}
\!\!\!\!\!A_\mu(x + L_j e_j + L_i e_i)&=& 
\Omega_j(x + L_i e_i) (A_\mu(x + L_i e_i)^* + \p_\mu) 
\Omega_j(x + L_i e_i)^\dagger \nonumber \\
&=&\Omega_j(x + L_i e_i) \Omega_i(x)^* (A_\mu(x) + \p_\mu) 
\Omega_i(x)^T \Omega_j(x + L_i e_i)^\dagger. \nonumber \\ \,
\end{eqnarray}
The two results are consistent only if the transition functions obey the first
cocycle condition of eq.(\ref{Ccocycle}). Next, we pick the spatial 
$i$-direction and the Euclidean time direction, such that
\begin{eqnarray}
\!\!\!\!\!A_\mu(x + L_i e_i + \beta e_4)&=&
\Omega_i(x + \beta e_4) (A_\mu(x + \beta e_4)^* + \p_\mu) 
\Omega_i(x + \beta e_4)^\dagger \nonumber \\
&=&\Omega_i(x + \beta e_4) \Omega_4(x)^* (A_\mu(x)^* + \p_\mu) 
\Omega_4(x)^T \Omega_i(x + \beta e_4)^\dagger. \nonumber \\ \,
\end{eqnarray}
Again, performing the two shifts in the opposite order we obtain
\begin{eqnarray}
\!\!\!\!\!A_\mu(x + \beta e_4 + L_i e_i)&=& 
\Omega_4(x + L_i e_i) (A_\mu(x + L_i e_i) + \p_\mu) 
\Omega_4(x + L_i e_i)^\dagger \nonumber \\
&=&\Omega_4(x + L_i e_i) \Omega_i(x) (A_\mu(x)^* + \p_\mu) 
\Omega_i(x)^\dagger \Omega_4(x + L_i e_i)^\dagger. \nonumber \\ \,
\end{eqnarray}
In this case, the resulting cocycle condition is the second one of 
eq.(\ref{Ccocycle}).

Eq.(\ref{Ctransition}) ensures the gauge-covariance of $C$-periodic boundary 
condition, i.e.
\begin{eqnarray}
A_\mu(x + L_i e_i)'&=&\Omega(x + L_i e_i) 
(A_\mu(x + L_i e_i) + \p_\mu) \Omega(x + L_i e_i)^\dagger 
\nonumber \\
&=&\Omega(x + L_i e_i) \Omega_i(x) (A_i(x)^* + \p_\mu) 
\Omega_i(x)^\dagger \Omega(x + L_i e_i)^\dagger \nonumber \\
&=&\Omega_i(x)' ({A_\mu(x)'}^* + \p_\mu) {\Omega_i(x)'}^\dagger.
\end{eqnarray}
Let us consider the gauge transformed cocycle condition
\begin{eqnarray}
&&\Omega_i(x + L_j e_j)' {\Omega_j(x)'}^* = 
\Omega_j(x + L_i e_i)' {\Omega_i(x)'}^* z_{ij}' \ \Rightarrow
\nonumber \\
&&\Omega(x + L_j e_j + L_i e_i)
\Omega_i(x + L_j e_j) \Omega(x + L_j e_j)^T 
\Omega(x + L_j e_j)^* \Omega_j(x)^* \Omega(x)^\dagger = 
\nonumber \\
&&\Omega(x + L_i e_i + L_j e_j) 
\Omega_j(x + L_i e_i) \Omega(x + L_i e_i)^T 
\Omega(x + L_i e_i)^* \Omega_i(x)^* \Omega(x)^\dagger z_{ij}' \
\Rightarrow \nonumber \\
&&\Omega_i(x + L_j e_j) \Omega_j(x)^* = 
\Omega_j(x + L_i e_i) \Omega_i(x)^* z_{ij}'.
\end{eqnarray}
Hence, $z_{ij}' = z_{ij}$, i.e. the twist-tensor is invariant under the
transformations of eq.(\ref{Ctransition}). Similarly, we obtain
\begin{eqnarray}
&&\Omega_i(x + \beta e_4)' {\Omega_4(x)'}^* = 
\Omega_4(x + L_i e_i)' \Omega_i(x)' z_{i4}' \ \Rightarrow
\nonumber \\
&&\Omega(x + \beta e_4 + L_i e_i)
\Omega_i(x + \beta e_4) \Omega(x + \beta e_4)^T 
\Omega(x + \beta e_4)^* \Omega_4(x)^* \Omega(x)^T = 
\nonumber \\
&&\Omega(x + L_i e_i + \beta e_4) 
\Omega_4(x + L_i e_i) \Omega(x + L_i e_i)^\dagger
\Omega(x + L_i e_i) \Omega_i(x) \Omega(x)^T z_{ij}' \
\Rightarrow \nonumber \\
&&\Omega_i(x + \beta e_4) \Omega_4(x)^* = 
\Omega_4(x + L_i e_i) \Omega_i(x) z_{i4}',
\end{eqnarray}
such that $z_{i4}' = z_{i4}$.

Interestingly, with $C$-periodic boundary conditions there are further
consistency conditions besides the cocycle condition eq.(\ref{Ccocycle}). For 
example, on the one hand, one has
\begin{eqnarray}
&&\Omega_i(x + L_j e_j + L_k e_k) \Omega_j(x + L_k e_k)^* \Omega_k(x) =
\nonumber \\ 
&&\Omega_j(x + L_i e_i + L_k e_k) \Omega_i(x + L_k e_k)^* \Omega_k(x) z_{ij} =
\nonumber \\
&&\Omega_j(x + L_i e_i + L_k e_k) \Omega_k(x + L_i e_i)^* \Omega_i(x) 
z_{ij} z_{ki} = \nonumber \\
&&\Omega_k(x + L_i e_i + L_j e_j) \Omega_j(x + L_i e_i)^* \Omega_i(x) 
z_{ij} z_{ki} z_{jk},
\end{eqnarray}
while, on the other hand,
\begin{eqnarray}
&&\Omega_i(x + L_j e_j + L_k e_k) \Omega_j(x + L_k e_k)^* \Omega_k(x) =
\nonumber \\ 
&&\Omega_i(x + L_j e_j + L_k e_k) \Omega_k(x + L_j e_j)^* \Omega_j(x) z_{kj} =
\nonumber \\
&&\Omega_k(x + L_j e_j + L_i e_i) \Omega_i(x + L_j e_j)^* \Omega_j(x) 
z_{kj} z_{ik} = \nonumber \\
&&\Omega_k(x + L_j e_j + L_i e_i) \Omega_j(x + L_i e_i)^* \Omega_i(x) 
z_{kj} z_{ik} z_{ji}.
\end{eqnarray}
Hence, unlike for periodic boundary conditions, there is the constraint of
eq.(\ref{Cmtwist}), $z_{ij}^2 z_{jk}^2 z_{ki}^2 = 1$, on the twist-tensor 
itself. Similarly, if one shifts in two spatial directions as well as in the 
Euclidean time direction, on the one hand, one finds
\begin{eqnarray}
&&\Omega_i(x + L_j e_j + \beta e_4) \Omega_j(x + \beta e_4)^* \Omega_4(x) =
\nonumber \\ 
&&\Omega_j(x + L_i e_i + \beta e_4) \Omega_i(x + \beta e_4)^* \Omega_4(x) 
z_{ij} = \nonumber \\
&&\Omega_j(x + L_i e_i + \beta e_4) \Omega_4(x + L_i e_i)^* \Omega_i(x)^* 
z_{ij} z_{i4}^* = \nonumber \\
&&\Omega_4(x + L_i e_i + L_j e_j) \Omega_j(x + L_i e_i) \Omega_i(x)^* 
z_{ij} z_{i4}^* z_{j4},
\end{eqnarray}
while, on the other hand,
\begin{eqnarray}
&&\Omega_i(x + L_j e_j + \beta e_4) \Omega_j(x + \beta e_4)^* \Omega_4(x) = 
\nonumber \\
&&\Omega_i(x + L_j e_j + \beta e_4) \Omega_4(x + L_j e_j)^* \Omega_j(x)^* 
z_{j4}^* =
\nonumber \\
&&\Omega_4(x + L_j e_j + L_i e_i) \Omega_i(x + L_j e_j) \Omega_j(x)^* 
z_{j4}^* z_{i4} = \nonumber \\
&&\Omega_4(x + L_j e_j + L_i e_i) \Omega_j(x + L_i e_i) \Omega_i(x)^* 
z_{j4}^* z_{i4} z_{ij}.
\end{eqnarray}
Consequently, one also obtains eq.(\ref{Cetwist}), $z_{i4}^2 = z_{j4}^2$.

\end{appendix}

\end{document}